\documentstyle[12pt]{article}
\def\vorepsilon{ {\overline\varepsilon} }
\def\vurepsilon{ {\underline\varepsilon} }

\def\be{\begin{equation}}
\def\fe{\end{equation}}

\def\spose#1{\hbox to 0pt{#1\hss}}\def\lta{\mathrel{\spose{\lower 3pt\hbox
{$\mathchar"218$}}\raise 2.0pt\hbox{$\mathchar"13C$}}}  \def\gta{\mathrel
{\spose{\lower 3pt\hbox{$\mathchar"218$}}\raise 2.0pt\hbox{$\mathchar"13E$}}} 

\def\Libra{\spose {--} {\cal L}}
\def\Euro{\spose {\lower 2.5pt\hbox{${^{\bf =}}$}}{ C}}
\def\ee{\end{equation}}

\def\bea{\begin{eqnarray}}
\def\eea{\end{eqnarray}}
\def\LLambda{  {\Lambda}}

\def\qq{ { q} }

\def\ggamma{ {\gamma}}

\begin{document}
\title{\bf Covariant Newtonian and Relativistic dynamics 
of (magneto)-elastic solid model for neutron star crust.}

\author { {\bf B. Carter$^\dag$, E. Chachoua$^\dag$, N. Chamel$^{\dag\star}$ }
\\ \hskip 1 cm\\   \\  \dag\ LuTh, Observatoire de Paris, 92195 Meudon, France,
\\ $\star$ Copernicus Astronomical Center (CAMK), 00-716 Warsaw, Poland.}

\date{\it 1 July, 2005}

\maketitle 

\vskip 1 cm

\noindent
{\bf Abstract.\ } 
This work develops the dynamics of perfectly elastic solid model for 
application to the outer crust of a magnetised neutron star. Particular 
attention is given to the Noether identities responsible for energy-momentum 
conservation, using a formulation that is fully covariant, not only (as is  
usual) in a fully relativistic treatment but also (sacrificing accuracy and 
elegance for economy of degrees of gravitational freedom) in the technically 
more complicated case of the Newtonian limit. The results are used to obtain 
explicit (relativistic and Newtonian) formulae for the propagation speeds of 
generalised (Alfven type) magneto-elastic perturbation modes.
 
\vskip 1.6 cm

\bigskip
{\bf 1. Introduction}
\medskip

In astrophysical contexts of the kind exemplified by a neutron star crust, it
is commonly necessary to go back and forth between relativistic models having
the advantage of greater elegance and in principle -- particularly at a global
level -- of higher accuracy, and Newtonian models that are more convenient 
from the point of view of other considerations such as economy in gravitational 
degrees of freedom, and availability of detailed underlying descriptions at a 
microscopic level. As a consequence of the fact that -- unlike the Galilean 
invariance group -- the Lorentz group is semi-simple, there are contexts (e.g. 
involving superfluidity~\cite{CK92,LSC97}) in which a fully relativistic 
treatment is actually easier to implement than a corresponding Newtonian 
treatment, even though the latter would be perfectly adequate as far as 
accuracy is concerned. On the other hand there are many contexts -- 
particularly those involving electromagnetic effects or strong gravitational 
fields -- in which a relativistic treatment would be indispensible if high
accuracy were required, but for which a Newtonian treatment might nevertheless 
be easier to implement as a first approximation. 

To facilitate transition between these two complementary kinds of description, 
it is desirable to develop a unified treatment in which both relativistic and 
Newtonian models are described in terms of technical machinery and terminology 
that are compatible as much as possible, so as to be consistent in the limit 
when the relativistic spacetime metric goes over to the degenerate spacetime 
structure of Newtonian theory. In a coherent approach of this kind, the 
Newtonian limit will naturally be obtained in fully covariant formulation of
the kind~\cite{Kunzle86,DuvalGibbonsHorvathy91,CK94} whose mathematical 
machinery was first developed by Cartan \cite{Cartan23}. In a preceding series 
of articles~\cite{CC02,CC03, CC04} on multiconstituent fluid and superfluid 
dynamics, it was shown how such a 4 dimensionally covariant formulation of 
Newtonian theory can provide physical insights (e.g concerning concepts such 
as helicity) that are not so easy to obtain by the traditional approach based 
on a 3+1 space time decomposition.

Continuing in the same spirit, the purpose of the present work is to 
contribute to the further development of the unified treatment of 
relativistic and Newtonian theory by treating the case of elastic solid 
models, of the kind appropriate for the description of the outer crust of a 
neutron star, including the magneto-elastic case (that arises in the limit of 
perfect electrical conductivity) for which the elastic structure is modified 
by a frozen-in magnetic field, of the kind whose effects are observed in 
pulsars. (The category of such models includes the special case of 
ordinary -- fluid not solid -- perfect  magnetohydrodynamics
in the limit of negligible elastic rigidity.)

Accurate description of such stars at a global level (not to mention a recently
proposed cosmological application~\cite{BCCM05}) requires a general 
relativistic treatment, but use of a flat space background will be sufficient
for treatment of the local mechanical properties to be considered here.
In ordinary pulsars the magnetic field is sufficiently low that (except in
the outer skin and the magnetosphere outside, where the matter density is 
comparatively low) such a flat background space treatment can be carried out 
(as described below) within a purely Newtonian framework.  However
a fully (at least special, if not general) relativistic treatment will be 
indispensible even locally (in a Minkowski background) when the magnetic field is 
sufficiently strong, as will be the case not just near the surface, but even in the 
deeper layers, for the special class of pulsars known as magnetars. The relativistic 
version of the magneto-elastic treatment developed here is particularly relevant for
such strongly magnetised  ($B\gta 10^{14} G$) neutron stars, in which flares 
powered by magnetic stress are believed to be responsible for 
gamma ray bursts of the brief but intense kind observed in soft gamma 
repeaters, the most spectacular example so far -- the most intense ever observed in
our galaxy -- having been the 27 December event that occurred in SGR $1806-20$ 
in 2004~\cite{Palmer2005}. For a complete treatment of such a flare a fully
general relativistic description would presumably be necessary since a phenomenon
of this kind is though to be attributable to a global modification of the magnetic
field of the neutron star~\cite{Eichler02,Hurley05}.

As an application, in both the Newtonian and fully relativistic cases such that
the underlying solid is in a simple isotropic state, the relevant (rigidity modified) 
propagation speeds of (Alfven type) perturbation modes are explicitly evaluated. 

A subsequent article will be needed to treat the more elaborate kind of 
model needed for the innermost layers of a neutron star crust, in which an ionic 
solid lattice is permeated by an independently moving current of superfluid
neutrons.

\bigskip
{\bf 2. Milne structure of Newtonian spacetime.}
\medskip

Before proceding, let us recapitulate the geometric essentials of 
Newtonian space time structure in a 4-dimensional background with respect 
to an arbitrary system of local coordinates $x^\mu$, $\mu=0,1,2,3$, as 
described in greater detail in the first article~\cite{CC02} of the 
preceding series.

The specification of a relativistic structure is fully determined by 
a non degenerate spacetime metric tensor having components $g_{\mu\nu}$, 
which in the special relativistic case are required to be constant in a 
prefered class of Minkowski type coordinate systems. This tensor will have 
a well defined contravariant inverse $g^{\mu\nu}$, from which a 
corresponding connection (which will vanish with respect to Minkowski 
coordinates in special relativity) with components 
$\Gamma_{\!\mu\ \rho}^{\ \nu}$ that are unambiguously obtainable using 
the usual Riemannian formula $\Gamma_{\!\mu\ \rho}^{\ \nu}=g^{\nu\sigma}
(g_{\sigma(\mu,\rho)}-{^1/_2}\,g_{\mu\rho,\sigma})$ using a comma to 
denote partial differentiation and round brackets to indicate index  
symmetrisation.

Newtonian theory is traditionally formulated in terms of an Aristotelian 
frame, meaning a direct product of a 1-dimensional trajectory 
parametrised by the Newtonian time $t$ and a flat 3-dimensional 
Euclidean space whose positive definite metric gives rise to a 
corresponding 4-dimensional metric $\eta_{\mu\nu}$ that is of degenerate, 
rank-3, positive indefinite type, so that it does not determine a
contravariant inverse tensor, and that, unlike its relativistic analogue, 
is not physically well defined because it depends on the choice of the 
Aristotelian ``ether'' frame, as characterised by the choice of a unit 
ether flow vector, $e^\mu$ say, that will be a null eigenvector of the 
corresponding degenerate metric, i.e. that will satisfy 
$\eta_{\mu\nu}e^\nu=0$.  There is however a complementary tensorial 
``Coriolis'' structure that (unlike the Aristotelian structure) is
physically well defined in the sense of being preserved by the allowable 
(time foliation preserving) ether gauge transformations. This invariant 
structure consists of the time gradient 1-form $t_\mu=t_{,\mu}$ and a 
contravariant metric tensor $\gamma^{\mu\nu}$ that (like its gauge 
dependent covariant complement $\eta_{\mu\nu}$)has the property of being 
degenerate, of rank-3 positive indefinite type, with null eigendirection 
determined by the time covector, i.e. {\be \gamma^{\mu\nu}t_\nu=0
\, .\label{1}\fe}

Although (like the non-degenerate metric in the relativistic case) this 
associated pair of tensors $t_\mu$ and $\gamma^{\mu\nu}$ is physically 
well defined, nevertheless the specification of this ``Coriolis'' 
structure (unlike that of the non-degenerate metric in the relativistic 
case) is not by itself sufficient to fully determine the geometric 
structure of spacetime in the Newtonian limit, and furthermore
(like the metric in special relativity but not in general relativity)
it is not freely variable over spacetime.

To start with, the Coriolis structure specified by the fields 
$\gamma^{\mu\nu}$ and $t_\mu$ is required to be {\it flat} in the sense 
that there should exist coordinates with respect to which the corresponding
field components are constant. It follows more particularly that there will 
be coordinates of a preferred type with respect to which these fields will 
have components of standard Aristotelian-Cartesian form, as given in terms 
of Kronecker notation by $\gamma^{\mu\nu}=\delta^\mu_{_1}\delta^\nu_{_1}
+\delta^\mu_{_2}\delta^\nu_{_2}+\delta^\mu_{_3}\delta^\nu_{_3}$, and
$t_\mu=\delta^{_0}_\mu$. Such a coordinate system will determine a 
corresponding Aristotelian ``ether'' frame vector with components 
$e^\mu=\delta^\mu_{_0}$ and its associated covariant space metric
with components $\eta_{\mu\nu}=\delta_\mu^{_1}\delta_\nu^{_1}+
\delta_\mu^{_2}\delta_\nu^{_2}+\delta_\mu^{_3}\delta_\nu^{_3}$, and
$t_\mu=\delta^{_0}_\mu$. (In the preceding work~\cite{CC02} this quantity 
$\eta_{\mu\nu}$ was written as $\gamma_{\mu\nu}$, using the same Greek 
letter gamma as for its contravariant  analogue, but in the present work 
the symbol $\gamma$ will be reserved for quantities that are gauge 
independent.) The Aristotelian-Cartesian coordinate system will also 
determine a corresponding symmetric connection, namely the one whose 
components
{\be  \Gamma_{\!\mu\ \rho}^{\ \nu}= \Gamma_{\!\rho\ \mu}^{\ \nu}
\label{1b}\fe}
 will vanish in that system, a requirement which evidently ensures that 
it will have vanishing curvature. The corresponding covariant
differentiation operator $\nabla$ will satisfy the commutation relation
{\be \nabla_{\!\rho}\nabla_{\!\sigma}-\nabla_{\!\sigma}\nabla_{\!\rho}=0
\, ,\label{2}\fe}
and will be such that the corresponding covariant derivatives of the 
tensor fields $\gamma^{\mu\nu}$ and $t_\mu$ will vanish:
{\be \nabla_{\!\rho}\gamma^{\mu\nu}=0\, \hskip  1 cm
 \nabla_{\!\rho}t_\mu=0\, .\label{3}\fe}
As well as satisfying  the algebraic conditions
{\be e^\mu t_\mu=1\, \hskip 1 cm e^\mu\eta_{\mu\nu}=0\, ,\hskip 1 cm 
\eta_{\mu\rho}\gamma^{\rho\nu}=\eta^\nu_\mu=
\delta_\mu^\nu-e^\nu t_\mu\, ,\label{4}\fe}
the ether frame dependent fields $e^\mu$ , $\eta_{\mu\nu}$, and the 
associated space projection tensor $\eta^\nu_\mu$  introduced in 
(\ref{4}), will also have corresponding covariant derivatives that vanish:
{\be \nabla_{\!\rho}e^\mu=0\, , \hskip  1 cm \nabla_{\!\rho}
\eta_{\mu\nu}=0\, \hskip 1 cm \nabla_{\!\rho}\eta^\nu_\mu=0
\,  .\label{5}\fe}

These fields can be used to specify an ether frame dependent Lorentz 
signature metric defined by $\overline g_{\mu\nu}=-t_\mu t_\nu+
\eta_{\mu\nu}$, with contravariant inverse, given by $\underline 
g{^{\mu\nu}}=-e^\mu e^\nu+\gamma^{\mu\nu}$, whose determinant provides 
a 4-dimensional volume measure that (modulo a sign ambiguity depending on 
a choice of parity orientation) fixes a corresponding antisymmetric 
tensor with components $\vorepsilon^{\mu\nu\rho\sigma}=
\vorepsilon^{[\mu\nu\rho\sigma]}$ (using square brackets to denote index 
antisymmetrisation). This measure tensor is alternatively definable 
directly by the condition that its components with respect to 
Aristotelian-Cartesian coordinates (with the chosen orientation) should 
be given by +1 or -1 whenever the indices are respectively even or odd 
permutations of the sequence $\{0,1,2,3\}$. The corresponding 
antisymmetric covariant measure 4-form $\vurepsilon_{\mu\nu\rho\sigma}=
\vurepsilon_{[\mu\nu\rho\sigma]}$ is then specifiable (in the manner that 
is familiar in the context of relativistic theory) by the normalisation 
condition
{\be \vorepsilon^{\mu\nu\rho\sigma}\vurepsilon_{\mu\nu\rho\sigma}=-4!
\  .\label{6}\fe}
It can be seen that (unlike $e^\mu$ and $\eta_{\mu\nu}$ and the 
frame dependent Lorentz metric ) they share with $t_\mu$ and 
$\gamma^{\mu\nu}$ the property of being independent of the
choice of ether gauge. These tensors will evidently give rise to
purely spacelike  3 index analogues, defined as the Hodge type duals of
$t_\mu$ and $e^\mu$ respectively, namely the gauge independent
space alternating tensor 
{\be  \epsilon^{\mu\nu\rho}=\vorepsilon^{\mu\nu\rho\sigma}t_\sigma
\, ,\label{7a}\fe}
and the frame dependent 3-form
{\be e^*_{\mu\nu\rho}=e^\lambda\vurepsilon_{\lambda\mu\nu\rho}
\label{7b}\fe}
which is interpretable as representing an ether current. It is evident 
that all these tensors will share the flatness property characterised by 
the connection, in the sense of satisfying the conditions
{\be \nabla_{\!\lambda}\vorepsilon^{\mu\nu\rho\sigma}=0\, ,\hskip 1 cm
\nabla_{\!\lambda}\vurepsilon_{\mu\nu\rho\sigma}=0
\, .\label{8}\fe}
and 
{\be \nabla_{\!\lambda}\epsilon^{\mu\nu\rho}=0\, ,\hskip 1 cm
\nabla_{\!\lambda}e^*_{\mu\nu\rho}=0
\, .\label{9}\fe}

Despite the simplification provided by the flatness property that 
is expressed by (\ref{3}), (\ref{5}), (\ref{8}) and (\ref{9}), the
Newtonian case is subject to the complication that neither the
flat coordinate system nor even the corresponding flat connection
is unambiguously determined by the tensor fields
$\gamma^{\mu\nu}$ and $t_\mu$. The standard form expressible by
$\gamma^{\mu\nu}=\delta^\mu_{_1}\delta^\nu_{_1}+
\delta^\mu_{_2}\delta^\nu_{_2}+\delta^\mu_{_3}\delta^\nu_{_3}$ and
$t_\mu=\delta^{_0}_\mu$ will in fact be preserved by a large category
of transformations that is known as the Coriolis group, which includes
not only boosts but also time dependent rotations. However the physical
structure of Newtonian spacetime is not preserved by time dependent
rotations, but only by transformations of a more restricted category
known as the Milne group, which is characterised by the condition
that the modification of the ether frame vector should depend only
on the Newtonian time $t$, as specified modulo a choice of origin
by $t_{,\mu}=t_\mu$. This means that the transformation of the
ether vector will be expressible in the form
{\be e^\nu\mapsto \breve e^\mu= e^\mu+b^\mu\, ,\label{10}\fe}
for a boost vector field $b^\mu$ that is subject to the condition
{\be b^\mu t_\mu=0\, ,\hskip 1 cm \gamma^{\nu\rho}\nabla_{\,\rho}b^\mu
=0\, \label{11}\fe}
and for which the corresponding acceleration vector will be 
specified by
{\be a^\mu=e^\rho\nabla_{\,\rho}b^\mu\, ,\hskip 1 cm a^\mu t_\mu=0
\, .\label{12}\fe}
The ensuing transformation of the covariant space metric will be 
given by
{\be \eta_{\mu\nu}\mapsto\breve\eta_{\mu\nu}=
\eta_{\mu\nu}-2t_{(\mu}\eta_{\nu)\rho}
b^\rho +\eta_{\rho\sigma}b^\rho b^\sigma t_\mu t_\nu\, ,\label{13}\fe}
while those of the corresponding space projection and
space measure tensors will be
{\be \eta^\nu_\mu\mapsto \eta^\nu_\mu-b^\nu t_\mu\, ,\hskip 1 cm
e^*_{\mu\nu\rho}\mapsto\breve e^*_{\mu\nu\rho}=
e^*_{\mu\nu\rho}+b^\lambda\vurepsilon_{\lambda\mu\nu\rho}
\, .\label{14}\fe}
Unlike the ether vector $e^\mu$, the covariant space metric
 $\eta_{\mu\nu}$, and the frame dependent Lorentz signature metric 
$g_{\mu\nu}$ that was invoked above, the tensors constituting the 
Coriolis structure, namely the time gradient covector $t_\mu$, the 
contravariant space metric $\gamma^{\mu\nu}$, and also the associated 
space-time measure given by $\vurepsilon_{\mu\nu\rho\sigma}$ or 
$\vorepsilon^{\mu\nu\rho\sigma}$ as well as the corresponding spacelike 
alternating tensor $\epsilon^{\mu\nu\rho}$ are all physically well 
defined in the sense of being independent of the choice of Aristotelian frame,
since the effect on them of the boost transformation specified by 
(\ref{10}) will be given trivially by
{\be \breve t_\mu=t_\mu\, \hskip 1 cm \breve\gamma^{\mu\nu}=
\gamma^{\mu\nu}\, ,\label{15}\fe}
and 
{\be \breve\vurepsilon_{\mu\nu\rho\sigma}=\vurepsilon_{\mu\nu\rho\sigma}\, ,
\hskip 1 cm \breve\vorepsilon{^{\mu\nu\rho\sigma}}=
\vorepsilon^{\mu\nu\rho\sigma}\, , \hskip 1 cm \breve\epsilon^{\mu\nu\rho}
=\epsilon^{\mu\nu\rho}\, .\label{16}\fe}

Within the full group constituted by the (in general non linear) Milne 
transformations characterised by (\ref{10}) and (\ref{11}), there is 
a linear subclass constituting the well known Galilei group, that is 
characterised by the requirement of preservation of the connection and 
the associated covariant differentiation operator $\nabla$, for which 
the necessary and sufficient condition is that the boost acceleration 
vector $a^\mu$ should vanish. However for a generic Milne transformation 
the covariant differentiation operator will undergo a non trivial 
transformation, $\nabla \mapsto \breve\nabla$, specified by a 
corresponding transformation of the connection that will be given,
using the definition (\ref{12}), by the formula
{\be \Gamma_{\!\mu\ \rho}^{\ \nu}\mapsto
\breve\Gamma_{\!\mu\ \rho}^{\ \nu}=\Gamma_{\!\mu\ \rho}^{\ \nu}
-t_\mu a^\nu t_\rho\, ,\hskip 1 cm 
\, ,\label{17}\fe} 
which has the noteworthy property of preserving the trace of the
connection, i.e. we shall have
{\be \breve\Gamma_{\!\mu\ \rho}^{\ \mu}=\Gamma_{\!\mu\ \rho}^{\ \mu}
\, ,\label{18}\fe}
with the implication that for the evaluation of the simple divergence
of a contravariant vector field, such as the displacement field
$\xi^\mu$ introduced below, it will not matter which connection we use, 
i.e. we shall have $\breve\nabla_{\!\mu}\xi^\mu=\nabla_{\!\mu}\xi^\mu$

Instead of working with the kind of flat but ether gauge dependent
connection that has just been described, it is useful for some purposes
to work instead with a curved but
gauge independent connection of a gravitational field dependent kind,
that was first introduced by Cartan, and that is described in the
preceding work referred to above~\cite{CC02}. However such a Cartan
connection will not be needed in the present work. 

\bigskip
{\bf 3. Relativistic correspondence}
\medskip

A Newtonian space time structure of the kind described in the preceding
section can be obtained as a low velocity limit from a corresponding
relativistic theory on the supposition that the latter is approximately 
describable, in terms of an adjustable parameter $c$, by a Lorentz 
signature metric 
${\rm d}\widetilde s^2=\widetilde g_{\mu\nu}\,{\rm d}x^\mu \,{\rm d}x^\nu$, 
having the form {\be \widetilde g _{\mu\nu}=\eta_{\mu\nu}-
\tilde c{^2}\,t_\mu\, t_\nu  \, ,\label{C1}\fe}
while, according to the preceding relations (\ref{4}), the corresponding
contravariant metric will be given by
{\be \widetilde g{^{\mu\nu}}=\gamma^{\mu\nu}-\frac{1}{\widetilde c{^2}}
\,e^\mu\, e^\nu \, .\label{C2}\fe}
The quantity $\tilde c$ in these expressions is interpretable as 
representing the speed of light with respect to coordinates of the 
standard Aristotelian kind as described in the preceding section. In the 
absence of gravitational effects the metric (\ref{C1}) can be taken to be 
of flat Minkowski type, as given by a fixed value for $\widetilde c$, but 
to allow for the effect of a Newtonian gravitational potential, $\phi$ 
say, it is necessary to take it to be given in terms of a fixed asymptotic 
value, $c$,  by the formula
{\be \widetilde c^2=c^2+2\phi\, .\label{C3}\fe}
The degenerate Newtonian structure of the preceding section is then 
obtained by taking the limit $ c\rightarrow\infty$, which evidently gives 
{\be \widetilde g{^{\mu\nu}}\rightarrow \gamma^{\mu\nu}\, ,\hskip 1 cm
c^{-2} \widetilde g_{\mu\nu}\rightarrow -t_\mu t_\nu\, .\label{C4}\fe}
It is to be remarked that although the spacetime metric itself is 
degenerate in this Newtonian limit, the associated Riemannian connection,
{\be \widetilde\Gamma_\mu{^\rho_{\ \nu}}=\widetilde g{^{\rho\sigma}}
(\widetilde g_{\sigma(\mu,\nu)}-\frac{_1}{^2}\widetilde g_{\mu\nu,\sigma})
\, ,\label{C5}\fe}
will be well behaved, and that it will agree with the flat connection
$\Gamma_\mu{^\rho_{\ \nu}}$ of the preceding section  in the 
absence of a gravitational field, i.e. when the potential $\phi$ in
(\ref{C3}) is uniform. There will however be a difference in the
generic case, for which it can be seen that the large $c$ limit will
be given by the relation
{\be \widetilde\Gamma_\mu{^\rho_{\ \nu}}\rightarrow \Gamma_\mu
{^\rho_{\ \nu}}+t_\mu t_\nu \gamma^{\rho\sigma}\phi_{,\sigma}\, .
\label{C6}\fe}
This shows that in the  Newtonian limit $\widetilde \Gamma_\mu
{^\rho_{\ \nu}}$ goes over to the Newton-Cartan connection that was 
denoted by $\omega_\mu{^\rho_{\ \nu}}$ in the preceding work~\cite{CC02}. 
This means that the associated Riemannian covariant differentiation 
operator $\widetilde\nabla_{\!\nu}$ will go over, not to the usual flat 
space differentiation operator $\nabla_{\!\nu}$ of the preceding section,
but to the Cartan type differentiation operator that was denoted in the 
preceding work~\cite{CC02} by $D_\nu$.

In Newtonian dynamical theory, the concept of mass conservation plays an 
essential role. In relativistic theory, mass as such is not in general
conserved, but in relevant applications it will nevertheless be possible to 
attribute most of the mass to other effectively conserved currents (e.g. 
that of baryons in a typical astrophysical context, or those of separate 
chemical elements in a typical non-nuclear terrestrial context). Such 
conserved currents can be endowed with suitable mass weighting factors  
-- e.g. the rest mass, $m$ say,  of an isolated proton or of a neutral  
hydrogen atom in the baryonic case -- so as to provide what is needed in a 
Newtonian limit. A conserved current can be represented --  without 
reference to any spacetime metric structure -- as a Cartan type 3-form, 
with antisymmetric components $n^*_{\mu\nu\rho}$ say, that is closed in 
the sense of having vanishing exterior derivative. This closure condition
will be equivalent to that of vanishing of the divergence of the 
corresponding current 4-vector that is given by the (Hodge type duality)  
ansatz
{\be n^\mu=\frac{c}{3!}\widetilde\varepsilon^{\mu\nu\rho\sigma}
\,n^\ast_{\nu\rho\sigma}\, ,\label{C7}\fe}
in terms of the antisymmetric measure tensor associated with the spacetime 
metric $\widetilde g_{\mu\nu}$, as given in terms of the modulus of the 
metric determinant $\vert \widetilde g\vert$ by the condition that its
components should be given by $+\Vert \widetilde g\Vert^{1/2}$ or 
$-\Vert \widetilde g\Vert^{1/2}$ whenever the indices are respectively
even or odd permutations of the sequence $\{0,1,2,3\}$. This means that it 
will be related to the non-relativistic spacetime measure of the 
preceding section by $\widetilde \varepsilon_{\mu\nu\rho\sigma}=
\widetilde c\,\vurepsilon_{\mu\nu\rho\sigma}$ and $\widetilde 
\varepsilon^{\mu\nu\rho\sigma} =\widetilde c^{-1}
\vorepsilon^{\mu\nu\rho\sigma}$, so that in the limit 
$c\rightarrow\infty$ one obtains $n^*_{\mu\nu\rho}\mapsto
\vurepsilon_{\mu\nu\rho\sigma}n^\sigma$).

A multiconstituent system may involve several constituent currents 
$n_{\rm _X}^{\,\nu}$, which need not be separately conserved, but that 
combine to give a locally conserved total
{\be n^{\nu}=\Sigma_{\rm _X}n_{\rm _X}^{\,\nu}\, ,\hskip
1 cm \widetilde \nabla_{\!\nu}n^{\nu}=0 \ .\label{C8}\fe}
For a confined system, the corresponding globally conserved mass integral,
$M$ say, associated with a spacelike hypersurface -- as specified by a fixed
value of some suitable time coordinate $x^{_0}$ -- will be given in terms
of the other coordinates and of the relevant mass parameter $m$ by
{\be M=m\int n^\ast_{_{1\, 2\, 3}}\,{\rm d}x^{_1}\,{\rm d}x^{_2}\,{\rm d}x^{_3}
\, ,\hskip 1 cm n^\ast_{_{1\, 2\, 3}}=c^{-1}\widetilde
\varepsilon_{_{1\, 2\, 3\, 0}}n^{_0} \, .\label{C9}\fe}

In the applications under consideration, the evolution equations of
the relevant currents will be obtainable from a  relativistic action 
principle of the world-line variational kind that is 
indispensible~\cite{C73} for treatment of a constituent that is solid, and 
that is very suitable~\cite{C89,CK92,LSC97} (though other -- e.g. Clebsch 
type~\cite{Schutz} -- possibilities exist) for treatment of a medium in 
which the relevant constituents are fluid.  This kind of variational 
principle is based on a relativistic action integral 
\be \widetilde {\cal I}=\int \widetilde\Lambda\,c^{-1}\widetilde
\varepsilon_{_{1\, 2\, 3\, 0}}\,{\rm d}x^{_1}
\,{\rm d}x^{_2}\,{\rm d}x^{_3}\,{\rm d}x^{_0} \, ,\label{C10}\fe 
for which the action density will be decomposible in the form 
{\be \widetilde\Lambda=\Lambda_{\rm_{bal}}+\Lambda_{\rm_{int}}
\, ,\label{C11}\fe}
in which $\Lambda_{\rm_{int}}$ is a relatively small intrinsic
contribution. The dominant extrinsic contribution $\Lambda_{\rm_{bal}}$ is
the ballistic part -- which is all that would be needed for the case of 
force free geodesic motion -- namely the negative of the sum of the 
relevant rest mass-energy contributions, as given by 
{\be \Lambda_{\rm_{bal}}=-m\, c^2\,\Sigma_{\rm _X}
n_{\rm _X}\, ,\label{C12}\fe} 
where $m$ is the appropriate mass weighting factor and, for each 
constituent, $n_{\rm _X}$ is the corresponding number density as 
evaluated in the relevant rest frame.  The  frame in question is 
characterised by the corresponding timelike frame vector 
$u_{\rm _X}^{\,\nu}$ that is specified, subject to the normalisation 
conditions $u_{\rm _X}^{\,\nu} u_{\rm _X\nu}=-c^2$, by the expressions
{\be n_{\rm _X}^{\,\nu}=n_{\rm _X}\, u_{\rm _X}^{\,\nu}\, ,
\hskip 1 cm n_{\rm _X}=c^{-1} \left( - n_{\rm _X}^{\,\nu}n_{\rm _X\nu}
\right)^{1/2}\, .\label{C13}\fe}

To obtain the Newtonian limit in which we are ultimately interested,
with respect to the gauge specified by some chosen ether vector
$e^\mu$, it is convenient to express each frame vector $u_{\rm _X}^{\,\nu}$ 
in terms of a corresponding purely spacelike 3-velocity vector 
$v_{\rm _X}^{\,\nu}$ in the form
{\be u_{\rm _X}^{\,\nu}=u_{\rm _X}^{\, _0}\left(e^\nu+v_{\rm _X}^{\,\nu}
\right) \, ,\hskip 1 cm v_{\rm _X}^{\,\nu}\, t_\nu=0\, ,\label{C14}\fe}
in which it can be seen that the required normalisation factor is
identifiable as the the time component of the frame vector with respect to
coordinates of the standard Aristotelian-Cartesian type  $x^{_1}=x,\ $ 
$x^{_2}=y,\ $ $x^{_3}=z,\ $ $x^{_0}=t, \ $ for which the metric takes 
the familiar form 
{\be {\rm d} \widetilde s^2 ={\rm d} x^2+{\rm d} y^2+{\rm d} z^2 
- \widetilde c^2 {\rm d} t^2\, .\label{C15}\fe}
This particular time component will evidently be expressible with respect  
to an arbitrary coordinate system by the covariant formula
{\be u_{\rm _X}^{\, _0}= u_{\rm _X}^{\,\nu}\, t_\nu= c\left(\widetilde c^2-
v_{\rm _X}^{\,2}\right)^{-1/2}\, ,\hskip 1 cm 
v_{\rm _X}^{\,2}= v_{\rm _X}^{\,\nu}\,v_{\rm _X\nu}=\eta_{\mu\nu}
 v_{\rm _X}^{\,\mu} v_{\rm _X}^{\,\nu}\, .\label{C16}\fe}
For the current itself we obtain the corresponding expression
{\be n_{\rm _X}^{\,\nu}=n_{\rm _X}^{\, _0}\left(e^\nu+v_{\rm _X}^{\,\nu}
\right) \, ,\hskip 1 cm  n_{\rm _X}^{\, _0}=  n_{\rm _X} u_{\rm _X}^{\, _0}=
n_{\rm _X}^{\,\nu}\, t_\nu\, ,\label{C17}\fe}
in which the relation between the rest frame number density $n_{\rm _X}$ 
and the corresponding ether frame component $n_{\rm _X}^{\, _0}$ will be 
given by
{\be n_{\rm _X}=n_{\rm _X}^{\, _0}\, c^{-1} \left(\widetilde c^2-
v_{\rm _X}^{\,2}\right)^{1/2}\, .\label{C18}\fe}

so that we shall have 
{\be n_{\rm_X}^{\, 2}-n_{\rm _X}^{\, _0\, 2}=n_{\rm _X}^{\, _0\, 2}
\left(\frac{2\phi}{c^2}-\frac{v_{\rm_X}^{\, 2}}{c^2}\right)\, .\label{C18a}\fe}
We thereby obtain a decomposition of the form
 {\be \Lambda_{\rm_{bal}}=\Lambda_{\rm_{ext}}+ \Lambda_{\rm_{rem}}\, ,\fe}
in which the extrinsic action contribution is given by
{\be \Lambda_{\rm_{ext}}=m\,\Sigma_{\rm_X}\frac{n_{\rm _X}^{\, _0\, 2}
(v_{\rm_X}^{\, 2}-2\phi)}
{ (n_{\rm_X}+n_{\rm _X}^{\, _0})}\, ,
\label{C18b}\fe}
which is evidently well behaved in the Newtonian limit, while the remaining 
contribution will be expressible in terms of the total current (\ref{C8}) in 
the simple form
 {\be \Lambda_{\rm_{rem}}= - m \,c^2 n^{_0}\, , \hskip 1 cm 
n^{_0}= n^\mu t_\mu\, .\label{C22}\fe}

Although this remainder $ \Lambda_{\rm_{rem}}$  will evidently be divergent 
in the large $c$ limit, it can be seen that this does not matter from the 
point of view of the variational principle since the corresponding 
integrated action contribution, as given in terms of the standard 
coordinates used for (\ref{C15}) by
{\be {\cal I}_{\rm _{rem}}=\int \Lambda_{\rm_{rem}}
\,c^{-1}\widetilde\varepsilon_{_{1\, 2\, 3\, 0}}\,{\rm d}x
\,{\rm d}y\,{\rm d}z\,{\rm d} t \, ,\label{C23}\fe} 
will be expressible in terms of the global mass function (\ref{C9}) 
simply as
{\be {\cal I}_{\rm _{rem}}=-\int M \,c^2\,{\rm d} t\, .\label{C24}\fe}
In the kind of application under consideration, the admissible variations
must respect the relevant current conservation law (\ref{C8}), so that 
they will have no effect on the global mass function $M$, which will 
therefore just be a constant that may be taken outside the integration. 
This means that the remainder term (\ref{C24}) will have a fixed value, so 
that from the point of view of the variation principle its inclusion is
entirely redundant. The ensuing theory is thus unaffected by replacing the 
original relativistic Lagrangian density $\widetilde\Lambda$ by a new but 
equivalent Lagrangian density 
{\be \Lambda_{\rm_{new}}=\widetilde \Lambda -\Lambda_{\rm_{rem}}
=\Lambda_{\rm_{ext}}+\Lambda_{\rm_{int}}
\, ,\label{C25}\fe}
from which the divergent term has simply been subtracted out. The new 
version  $\Lambda_{\rm_{new}}$ has the technical disadvantage of being 
gauge dependent, since its specification depends on the choice of the 
ether frame characterised by the vector $e^\mu$, but it  has the important 
advantage of remaining well behaved  in the Newtonian limit. With the usual 
assumption that (with respect to the chosen ether frame) the relevant 
squared space velocities and the potential are of the same order of 
smallness compared with the speed of light, we shall obtain
{\be n_{\rm _X}-n_{\rm _X}^{\, _0}= n_{\rm _X}\left(
\frac{\phi}{c^2}-\frac {v_{\rm _X}^{\,2}}{2 c^2}\right)
+ {\cal O}\left\{\frac {v^4}{c^4}\right\}\, ,\label{C19}\fe}
as $c\rightarrow\infty$. It can thereby be seen that in this limit the new 
version of the action density will take the form
{\be \Lambda_{\rm_{new}}= \Lambda+ 
{\cal O}\left\{\frac {v^4}{c^2}\right\}\, ,\label{C26}\fe}
in which $\Lambda$ is of purely Newtonian type, taking the standard 
generic form that was postulated at the outset in the preceding 
work~\cite{CC02,CC03},  namely
{\be \Lambda=\Lambda_{\rm_{int}}+\Lambda_{\rm_{ext}}\, ,\hskip 1 cm
\Lambda_{\rm_{ext}}=\Lambda_{\rm_{kin}}+\Lambda_{\rm_{pot}}\, ,\label{C27}\fe}
in which the kinetic and gravitational potential contributions have their 
usual Newtonian form
{\be \Lambda_{\rm_{kin}}= \frac{_1}{^2} m\,\Sigma_{\rm _X} n_{\rm _X} 
v_{\rm _X}^{\,2}\, ,\hskip 1 cm \Lambda_{\rm_{pot}}
=-\phi \,m \,\Sigma_{\rm _X} n_{\rm _X}\, ,\label{C21}\fe}

\bigskip
{\bf 4. Noetherian construction of stress - energy tensor.}
\medskip

For a variationally formulated theory in a general relativistic
context, the corresponding stress momentum energy tensor is commonly 
constructed directly by differentiation of the relevant action density 
with respect to the spacetime metric. The purpose of the present section 
is to describe the adaptation of such a procedure to the technically 
more complicated Newtonian case.

In the preceding work on fluid systems~\cite{C89,CC02,CC03} it was 
convenient to formulate the action in terms of physical fields (such as 
current 4-vectors) that are not entirely free but partially restrained as 
far as the application of the variation principle is concerned. However in 
the case of the solid systems with which  the present article will be 
concerned it will be more convenient (albeit at the expense of greater 
gauge dependence) to work just with fields whose variation is unrestrained. 
The advantage of using an unrestrained variational formulation as we shall 
do here is that for an unperturbed field configuration satisfying the 
dynamical equations provided by the variational principle, the most 
general variation of the action will be unaffected, modulo the addition
of a variationally irrelevant divergence, by the variations of the 
relevant dynamical fields, and will therefore be given, modulo such 
a divergence, just by the contributions from variations of the given  
background fields characterising the relevant Newtonian or relativistic 
spacetime structure.

In the relativistic case, the only independently given background field
needed for this purpose is the metric $\tilde g_{\mu\nu}$ itself.
Provided that the other dynamical fields in the Lagrangian 
$\widetilde \Lambda$  obey the
corresponding variational field equations, the generic action variation
will be given by an expression of the simple form
{\be \delta\widetilde\Lambda\cong \frac{\partial\widetilde\Lambda}
{\partial \widetilde g_{\mu\nu}}\,\delta \widetilde g_{\mu\nu} \, .
\label{C28}\fe}
using the symbol $\cong$ to denote equivalence modulo a divergence, 
i.e. modulo a term that is variationally irrelevant because its
integral for a perturbation in a confined spacetime domain will
vanish by Green's theorem.

As has long been well known in the context of general relativistic 
theory~\cite{Trautman65}, and as has more recently been demonstrated 
in the Newtonian case~\cite{CC03}, the use of a fully covariant formulation
makes it possible to derive useful Noether type identities by considering
variations of the trivial kind generated by an arbitrary displacement
field, $\xi^\mu$ say. This means that the
variation of each (background or dynamical) field variable will be given by
the negative of its Lie derivative. In the relativistic case,
the relevant variations will be given simply by
{\be -\delta\widetilde\Lambda=
{\vec \xi}\Libra\,\widetilde\Lambda\equiv\xi^\nu\widetilde\nabla_{\!\nu}
\widetilde\Lambda\cong -\widetilde\Lambda\widetilde\nabla_{\!\nu} 
\xi^\nu\, , \label{C29}\fe}
and 
{\be -\delta \tilde g_{\mu\nu}={\vec \xi}\Libra\,\widetilde g_{\mu\nu}
=2\widetilde\nabla_{\!(\mu}\xi_{\nu)}\, .\label{C30}\fe}
Their substitution in (\ref{C28}) provides a relation of the form
{\be \widetilde T^{\mu\nu}\widetilde\nabla_{\!(\mu}\xi_{\nu)}\cong0\, 
, \label{C31}\fe}
in which the relevant stress momentum energy density tensor can be
read out as
{\be  \widetilde T^{\mu\nu}=2\frac{\partial\widetilde\Lambda}
{\partial \widetilde g_{\mu\nu}}+\widetilde \Lambda\,
\widetilde g^{\mu\nu}\, .\label{C32}\fe}
By again removing a variationally irrelevant divergence,  (\ref{C31}) 
can be rewritten equivalently as
{\be \xi^\mu\widetilde\nabla_{\!\nu}\widetilde T{^\nu}_{\!\mu}
\cong 0\, ,\label{C33}\fe
}
which, since it must hold for a locally arbitrary vector field
$\xi^\nu$, shows that the variational field equations entail -- as a 
generic consequence -- a divergence condition of the well known form
{\be \widetilde\nabla_{\!\nu}\widetilde T{^\nu}_{\!\mu}
=0\, ,\label{C34}\fe}
which in the absence of gravity, i.e. in a flat Minkowski background,
is interpretable as an energy-momentum conservation law.

The (automatically symmetric) ``geometric'' kind of stress-energy 
tensor (\ref{C32}) needs to be distinguished from the (related, but in 
general different and not necessarily symmetric) kind of stress-energy 
tensor commonly referred to as ``canonical'', which is constructed by 
differentiation not with respect to the metric, but with respect to the 
other dynamically relevant fields. Even in a special relativistic context, 
i.e. when gravitation is negligible  so that the physical metric can be 
taken to be fixed, it is still perfectly legitimate (though the possibility 
of doing so is not widely realised) to exploit the greater convenience of 
the ``geometric'' construction via the consideration of virtual (``off 
shell'') variations of the metric. However the construction of a
``geometric'' stress energy tensor is not quite so straightforward in a
non-relativistic Newtonian framework, due to the degeneracy of the metric, 
which makes it harder to avoid the inelegancies of the traditional 
``canonical'' treatment. 

Although it is not quite so simple and convenient as in the relativistic
case, a ``geometric'' type ansatz for the construction of an appropriate
 variational stress-energy tensor can nevertheless be obtained in
a Newtonian framework using the 4-dimensionally covariant formalism 
set up in the previous section. Such an ansatz was developed in
the preceding work~\cite{CC03}, where it was shown how, in the case of
a simple or multiconstituent fluid model, the action density function
$\Lambda$ for a 4-dimensionally covariant variational formulation provides 
a Noether identity that leads automatically to a natural ``geometric''
type ansatz for a correponding non-relativistic stress-momentum-energy 
density tensor $T{^\mu}_{\!\nu}$.  The present section describes the
way to obtain the appropriate Noetherian ansatz for the non-relativistic 
``geometric'' stress-energy tensor in a manner that is rather simpler than 
was required for the partially restrained case dealt with in the 
preceeding work~\cite{CC03}.

The kind of system under consideration is one governed
by a non relativistic  action integral of the form
\be {\cal I}=\int \Lambda\,\vurepsilon_{_{1\, 2\, 3\, 0}}\,{\rm d}x^{_1}
\,{\rm d}x^{_2}\,{\rm d}x^{_3}\,{\rm d}x^{_0} \, ,\label{19}\fe 
in which the action density $\Lambda$ is a scalar of the generic form
(\ref{27}) that is formulated as a function just of the relevant 
(variationally unrestrained) dynamical fields and their gradients, and of 
the background fields $t_\mu$ and $\gamma^{\mu\nu}$ characterising the 
Milne structure of Newtonian spacetime, as well as on a gauge dependent ether
frame vector field $e^\mu$. The involvement of the latter will violate 
the Milne and even Galileian invariance of the local action density, but 
will not violate the required covariance of the ensuing field equations 
provided the effect of an ether gauge transformation $\Lambda\mapsto
\breve\Lambda$ is just to add on a pure divergence term, which will not 
contribute to the integral of a local variation. This requirement is 
conveniently expressible as $\breve\Lambda\cong\Lambda$, again using the 
symbol $\cong$ to denote equivalence modulo a (variationally irrelevant) 
divergence, for whose construction (due to the ether frame invariance 
of the measure tensor $\vurepsilon_{\mu\nu\rho\sigma}$ as remarked at 
the end of the preceding section) it makes no difference whether we use 
the original covariant differentiation operator, $\nabla$, or the modified 
operator $\breve\nabla$.

As remarked above, the simplification in the unrestrained case dealt with 
here is that for an unperturbed field configuration satisfying the 
dynamical equations provided by the variational principle, the most general 
variation of the action will be unaffected, modulo the addition  of a 
variationally irrelevant divergence, by the variations of the relevant 
dynamical fields. Modulo such a divergence, the local action variation 
will therefore be given, in the non-relativistic case, just by the 
contributions from variations of the uniform background fields 
$\gamma^{\mu\nu}$, $t_\mu$ and $e^\mu$, as well as of a generically 
non-uniform gravitational potential $\phi$ in cases for which the latter 
is introduced as given non dynamical background. The variation will 
therefore be given, modulo a divergence, in terms just of a set of 
tensorial coefficients $\partial\Lambda/\partial\gamma^{\mu\nu}$,
$\partial\Lambda/\partial t_\mu$, and $\partial\Lambda/\partial e^\mu$ 
(of which the latter would not be needed in a model with ether gauge 
independent action density) by an expression of the form
\be \delta\Lambda\cong{\partial\Lambda\over\partial\gamma^{\mu\nu}}
\,\delta\gamma^{\mu\nu}+{\partial\Lambda\over\partial t_\mu}\,\delta t_\mu
+{\partial\Lambda\over\partial e^\mu}\,\delta e^\mu
+{\partial\Lambda\over\partial\phi}\delta \phi
\, ,\label{20}\fe
It is to be remarked that this expression is not by itself sufficient
to fully determine the coefficients involved. Further suitably chosen
conventions, of which the most obviously appropriate is that the tensor
$\partial\Lambda/\partial\gamma^{\mu\nu}$ should be symmetric, in
view of the algebraic restrictions on the independence of the
variations involved, which by (\ref{1}) and (\ref{4}) must must 
evidently satisfy
\be t_\mu\delta\gamma^{\mu\nu}=-\gamma^{\mu\nu}\delta t_\mu\, ,
\hskip 1 cm t_\mu\delta e^\mu=- e^\mu\delta t_\mu\, .\label{21}\fe
 For the actual evaluation of the coefficients in (\ref{20}) it will
also be useful to have formulae for the variations of other 
associated spacetime background fields that may be involved in the
explicit formulation of the action, such as the covariant spacetime
metric whose variation will be given by the formula
\be \delta\eta_{\mu\nu}=-\eta_{\mu\rho}\eta_{\nu\sigma}\delta
\gamma^{\rho\sigma}-2t_{(\mu}\eta_{\nu)\lambda}\delta e^\lambda
\, ,\label{22}\fe
and the spacetime alternating tensor whose variation will be given by
the formula
\be \delta\vorepsilon^{\mu\nu\rho\sigma}=\vorepsilon^{\mu\nu\rho\sigma}
 ({_1\over^2}\eta_{\kappa\lambda}\delta \gamma^{\kappa\lambda}
-e^\lambda\delta t_\lambda)\, ,\label{23}\fe
(in which the ether frame dependence of the two separate terms can
be seen to cancel out to give a gauge invariant total).

As in the relativistic case above, we now consider the effect of variations 
of the trivial kind generated by an arbitrary displacement field, $\xi^\mu$ 
say, so that the variation of each (background or dynamical) field variable 
will be given by the negative of its Lie derivative. The relevant formulae 
are thus given by
\be -\delta\Lambda={\vec \xi}\Libra\,\Lambda\equiv\xi^\rho\nabla_{\!\rho}
\Lambda\, , \label{24}\fe
\be -\delta\gamma^{\mu\nu}={\vec\xi}\Libra\,\gamma^{\mu\nu}\equiv
\xi^\rho\nabla_{\!\rho}\gamma^{\mu\nu}-2\gamma^{\rho\,(\mu}
\nabla_{\!\rho}\xi^{\nu)}\, ,\label{25}\fe
\be -\delta t_\mu={\vec\xi}\Libra\, t_\mu\equiv
\xi^\rho\nabla_{\!\rho}t_\mu+t_\rho\nabla_{\!\mu}\xi^\rho\, ,\label{26}\fe
\be -\delta e^\mu={\vec\xi}\Libra\, e^{\mu}\equiv
\xi^\rho\nabla_{\!\rho}e^\mu-e^\rho\nabla_{\!\rho}\xi^\mu\, .\label{27}\fe
\be -\delta\phi={\vec \xi}\Libra\,\phi\equiv\xi^\rho\nabla_{\!\rho}
\phi\, , \label{24a}\fe

The first of these formulae can be rewritten, modulo a divergence, as
\be \delta\Lambda\cong\Lambda\nabla_{\!\rho}\xi^\rho\, ,\label{28}\fe
while in view of the uniformity properties (\ref{3}) and (\ref{5}) of the 
unperturbed background fields, the next three will reduce to the form
\be \delta\gamma^{\mu\nu}=2\gamma^{\rho\,(\mu}
\nabla_{\!\rho}\xi^{\nu)}\, ,\label{29}\fe
\be \delta t_\mu=-t_\rho\nabla_{\!\mu}\xi^\rho\, ,\label{30}\fe
\be \delta e^\mu=e^\rho\nabla_{\!\rho}\xi^\mu\, .\label{31}\fe
For such a displacement variation, the relation (\ref{20}) will
therefore reduce to the form
\be T^\mu_{\ \nu}\nabla_{\!\mu}\xi^\nu\cong \rho\,\xi^\nu\nabla_{\!\nu}\phi
\, ,\label{32}\fe
in terms of a stress-momentum energy density tensor  $T_{\ \nu}^\mu$
a gravitational mass density $\rho$ that can be read out as
\be T_{\ \nu}^\mu=\Lambda\delta^\mu_{\,\nu}-
2{\partial\Lambda\over\partial\gamma^{\rho\nu}}\,\gamma^{\rho\mu}
+ {\partial\Lambda\over \partial t_\mu}\, t_\nu -{\partial
\Lambda\over\partial e^\nu}\, e^\mu\, ,\label{33}\fe
and
\be \rho=-{\partial\Lambda\over\partial \phi}\, .\label{33a}\fe

By a further divergence adjustment the equivalence (\ref{32}) can be
rewritten as
\be \xi^\nu(\nabla_{\!\mu}T^\mu_{\ \nu}+\rho\nabla_{\!\nu}\phi)
\cong 0\, ,\label{34}\fe
which means that for a displacement confined to a localised
spacetime region outside which the hypersurface contribution provided
via Green's theorem by the unspecified divergence term will vanish)
we shall have 
\be\int  \xi^\nu(\nabla_{\!\mu}T^\mu_{\ \nu}+\rho\nabla_{\!\nu}\phi)
\,\varepsilon_{_{0\, 1\, 2\, 3}}\,{\rm d}x^{_0}
\,{\rm d}x^{_1}\,{\rm d}x^{_2}\,{\rm d}x^{_3}=0\, .\label{35}\fe 
Since this identity must hold for an arbitrarily chosen displacement
field $\xi^\mu$ in the spacetime region under consideration, it follows
that we must have 
 \be \nabla_{\!\mu}T^\mu_{\ \nu}=-\rho\nabla_{\!\nu}\phi\, ,\label{36}\fe
as a Noether type identity. We have thus established a theorem
to the effect that the conservation law (\ref{36}) will hold automatically
for the geometric energy momentum tensor obtained from the 
covariantly formulated Newtonian action density $\Lambda$ 
according to the prescription (\ref{33}) whenever
the dynamical field equations provided by the corresponding
unrestrained variation principle are satisfied.

In the simple applications to be considered below in the present article,
the local energy momentum conservation law (\ref{36}) will, by itself
amount to complete set of dynamical field equations, but it will
only be a subset theroff in more general cases, such as the
multiconstituent applications that we plan to deal with in subsequent work.

\bigskip
{\bf 5. The Material Projection concept}
\medskip

The historical development of the standard textbook theory (see e.g. 
Landau and Lifshitz~\cite{LL}) of a perfectly elastic solid in a 
Newtonian context is attributable to many people, among whom Cauchy is 
perhaps the most notable. However, as in the multiconstituent fluid 
case~\cite{CK94,CC02}, the insight needed for the formulation of a generally
covariant version of the theory comes rather from its relativistic 
generalisation, for which a fully covariant formulation has always been 
used, as an indispensible necessity, from the outset. Some of the earliest
work on the appropriate relativistic theory of a simple perfect solid was 
carried out in a purely mathematical context by Souriau~\cite{Souriau65}, 
and by DeWitt~\cite{DeWitt62} (who needed it as a toy model for testing
techniques to be used in the quantisation of gravity). Its development as 
a realistic physical theory for use in the kind of astrophysical context
(particular that of neutron stars) that motivates the present work was 
initiated  (in the aftermath of the discovery of pulsars) with Quintana
by one of the present authors~\cite{CQ72}, in a formulation~\cite{C83} 
that was shown to be elegantly obtainable by a variational 
approach~\cite{C73,C89} that will be used as a guide for the present work, 
whose purpose is to derive its Newtonian analogue.

The material projection is based on the simple consideration that the 
intrinsic structure of a solid is essentilly 3 dimensional. This means
that the mathematical entities (such as differential forms) that will 
be used in a variational principle governing the dynamic behaviour
of the solid should be defined over a 3 dimensional  manifold, $\cal{X}$ 
say. The prototypical example of such field is the elastic-stress tensor, 
whose definition should not depend  on  the solid's history, whereas
its explicit value obviously will do so. The requisite 3 manifold $\cal{X}$
is identifiable  as the quotient of spacetime by the worldines of the 
idealised particles (representing microscopic lattice sites)
constituing the solid , so that each point on $\cal{X}$  can be considered 
as the projection of the world-line describing the extrinsic motion of the 
relevant particle. Thus a patch of local coordinates, let us say $q^{_A}$ 
(for $A$=1,2,3), on ${\cal X}$ will induce a corresponding set of scalar 
fields $q^{_A}$ that will be given as functions of the local spacetime 
coordinates, $x^\mu$ (for $\mu$ = 0, 1, 2, 3), on ${\cal M}$. 

The local scalar fields $q^{_A}$ can be interpreted as a set of comoving 
-- Lagrange type -- coordinates on spacetime. They might even be used to 
specify the choice of the space coordinates by taking $x^{_1}=q^{_1}$, 
$x^{_2}=q^{_2}$ $x^{_3}=q^{_3}$. However such a choice is more likely to 
be convenient in a general relativistic context than in that of a flat 
Minkowski background, or in the Newtonian case, for which a more commonly 
convenient choice is to use background coordinates $x^\mu$ flat 
(respectively Minkowski or Aristotelian-Cartesian) in order to simplify 
the procedure of covariant differentiation (in a manner that is not 
possible in the general relativistic case) by setting the connection 
coefficients $\Gamma_{\!\mu\ \rho}^{\ \nu}$ everywhere to zero.

Since the whole worldline  of the particle is mapped into a point
in ${\cal X}$ the $\qq^A$, when viewed as scalar fields on $\cal{M}$, will
obviously be characterised by the  property
{\be {\vec u}\Libra q^{_A} = u^\mu \nabla_\mu q^{_A} =u^\mu q^{_A}_{\, ,\mu}=0 
\, ,\label{E0}\fe}
where the tangent vector field $u^\mu$ to the worldline is subject
to the standard normalisation condition given by $u^\mu u_\mu=-c^2$ in the 
relativistic case, and hence by $u^\mu t_\mu=1$ in the Newtonian limit 
(\ref{C4}). Using the symmetry property (\ref{1b}) of the connection, which 
ensures that we shall have $\nabla_\mu q^{_A}_{\, ,\nu}=
\nabla_\nu q^{_A}_{\, ,\mu}$, the relation (\ref{E0}) in turns implies that
{\be {\vec u} \Libra q^{_A}_{\, ,\mu}=u^\nu\nabla_\nu q^{_A}_{\, ,\mu}
+q^{_A}_{\, ,\nu}\nabla_\mu u^\nu =0 \, . \label{E0bis}\fe}

Let us now consider the example of a material 1-form on the manifold $\cal X$,
with components $A_{_B}$ say. When pulled back via the material projection,  
this material form induces a spacetime covariant 1-form on the spacetime 
manifold $\cal M$ with components $A_\mu$ given simply by
\be
A_{\mu}=A_{_B} q^{_B}_{\, ,\mu}\, ,
\label{pullback}
\ee
with the direct implication that one will have
\be
A_{\mu} u^{\mu}=0 \, , ~~~~~~  {\vec u} \Libra A_\mu =2 u^\nu 
\nabla_{[\nu} A_{\mu]} = 0 \label{f}
\ee
for any such material 1-form. Conversely if any spacetime 1-form is such that
it satisfies both of the conditions above, then it is
uniquely determined by a material 1-form through the pullback operation.

The generic defining property of the kind of simple perfectly elastic
model to be considered here is that the action should depend only
on the rheometric position coordinates $q^{_A}$ and on the corresponding
induced contravariant metric components $\gamma_{_{AB}}$,  or 
equivalently on the corresponding contravariant components 
$\gamma^{_{AB}}$, which are defineable by the reciprocity relation 
 \be
\gamma^{_{AB}}\gamma_{_{BC}}=\delta^{_A}_{\,_C} \, ,\label{60}
\fe
and which will be given in the relativistic case simply as the 
rheometric projection of the contravariant spacetime metric,
\be
 \gamma^{_{AB}}=\widetilde g^{\mu\nu}q^{_A}_{\, ,\mu}q^{_B}_{\, ,\nu}
\, ,\label{E2}
\fe
while in the Newtonian limit (\ref{C4}) the required components
 will be given by 
\be 
\gamma^{_{AB}}=\gamma^{\mu\nu}q^{_A}_{\, ,\mu}q^{_B}_{\, ,\nu}
\, .\label{47}
\fe

It is to be remarked that, in solid models of the kind appropriate for 
typical (low pressure) laboratory type terrestrial applications, the 
rest-frame energy per particle  will commonly have an absolute minimum for 
some preferred value,  $\kappa_{_{AB}}$ say, of the induced metric 
$\gamma_{_{AB}}$. In such such cases  $\kappa_{_{AB}}$ can be considered, 
and exploited, as a natural fixed (positive definite) Riemannian metric on the
material base space. In other words  
$\kappa_{_{AB}} {\rm d }q^{_A}{\rm d }q^{_B}$ represents the ``would be"  
relaxed distance between the chosen particle and a nearby one in the 
frame of the former, assuming the attainability of such relaxed state (which
could be defined by extracting a piece of the continuous medium --
i.e. a neighbourhood of the considered particle -- and
letting it reach the minimum local energy density state). In generic
circumstances however, such a preferred rheometric metric may be
 ill defined, since a local state of minimised energy need not 
exist. This caveat applies particular in cases of very high pressure (such 
as will occur in deep stellar interiors) from which a process relaxation 
might lead, not to a minimised energy state for the solid, but merely to 
its vaporisation as a gas.

The formula (\ref{47}), while defined in the Newtonian limit,  will also be 
valid in the relativistic case provided it is interpreted in terms of the 
relevant rank-3 worldline-tangential projection tensor, which will be given by
{\be \gamma^\mu_{\,\nu}=\delta^\mu_{\nu}+c^{-2}u^\mu u_\nu
\, ,\label{E4}\fe}
in the relativistic case, and which will will go over,
in the Newtonian limit (\ref{C4}),  to
{\be \gamma^\mu_{\, \nu}=\delta^\mu_{\nu}-u^\mu t_\nu
\, .\label{E5}\fe}
(Except in the case of a static configuration -- for which the ether 
vector $e^\mu$ may be chosen to coincide with $u^\mu$ --  the variable 
flow-tangential projection tensor defined by (\ref{E5})  must be 
distinguished from the uniform ether frame dependent projection 
tensor (\ref{4}) that was denoted by the same symbol in the preceding 
work~\cite{CC02,CC03} but that is denoted here by $\eta^\mu_{\,\nu}$).

We shall use the induced base metric of (\ref{60}) for raising and lowering 
of the material base indices in the usual way, as illustrated for a
covector with components $A_{_A}$ by
{\be A^{_A}=\gamma^{_{AB}}A_{_B}\, ,\hskip 1 cm 
A_{_A}=\gamma_{_{AB}}A^{_B}\, ,\label{61}\fe}

The (variable) covariant metric as defined by 
\be \gamma_{\mu\nu}= \gamma_{_{AB}}q^{_A}_{\, ,\mu}q^{_B}_{\, ,\nu}
\, .\label{E7}
\fe
 can be used in conjunction with the (uniform) contravariant metric 
$\gamma^{\mu\nu}$
for an unambiguously reversible index raising lowering and raising
operation for vectors and tensors that are orthogonal to the material 
flow in the manner exemplified, according to (\ref{f}),
by the pull back (\ref{pullback}) of $A_{_A}$ namely
\be A_\mu=A_{_A}q^{_A}_{\, \mu}=\gamma_{\mu\nu}A^\nu\, ,\label{68}\fe
whose raised version will project onto $A^{_A}$:
\be A^\mu=\gamma^{\mu\nu}A_\nu\, ,\hskip 1 cm A^\mu t_\mu=0\, ,\hskip 
1 cm q^{_A}_{\, \mu} A^\mu= A^{_A}\, .\label{69}\fe

Is to be noted that although the procedure defined by (\ref{61}) is 
invariant with respect to the choice of ether frame, it does depend on 
time. Thus if $A_{_A}$ is taken to be a fixed vector potential 
characterising the ``frozen-in'' magnetic field that will be introduced 
below, so that its time derivative $\dot A_{_A}$ will vanish, the 
corresponding contravariant vector will neverthess be time dependent:
\be
 \dot A_{_A}=0\, \ \Rightarrow\ \dot A^{_A}=\dot \gamma^{_{AB}}
A_{_B}\, .\label{62}
\fe

It can be seen from (\ref{3}) and (\ref{E0bis}) that the  time
derivative, along the worldlines, of the induced base metric will be given by
{\be \dot\gamma_{_{AB}}=2\theta_{_{AB}}=
-\gamma_{_{AC}}\gamma_{_{BD}}\dot\gamma^{_{CD}}\, ,
\hskip 1 cm \dot \gamma^{_{AB}}=-2\theta^{_{AB}}=
-2 q^{_A}_{\, ,\mu}q^{_B}_{\, ,\nu}\theta^{\mu\nu}\, ,\label{64}\fe}
in terms of the (symmetric) strain rate tensor
{\be \theta^{\mu\nu}=\gamma^{\rho(\mu}\nabla_{\!\rho}u^{\nu)}\,
.\label{65}\fe}
which will automatically  satisfy the orthogonality
condition $\theta^{\mu\nu}u_\nu=0$ in the relativistic case, so that
we shall have $\theta^{\mu\nu}t_\nu=0$ in the Newtonian limit.

The time derivation considered so far has concerned only quantities
such as the base space components that have the status of scalars
from the point of view of the background space time. We now extend the
dot notation to quantities that are tensorial with respect to the
backkground  spacetime by defining it to indicate covarant differentiation
with respect to time along the world lines, meaning that $\dot{\,}$ will
indicate the effect of the operator $u^\rho\nabla_{\!\rho}$ in the
manner illustrated by the definition of the acceleration vector which
will be given by 
\be \dot u^\mu=u^\rho\nabla_{\!\rho} u^\mu\, .\label{70}\fe 
so that it will satisfy the condition $\dot u^\mu u_\mu=0$ in the 
relativistic case and $\dot u^\mu t_\mu=0$ in the Newtonian limit.
It can be seen that the corresponding covariant time derivatives 
$\dot\gamma_{\mu\nu}\equiv u^\rho\nabla_{\!\rho}\gamma_{\mu\nu}$
of the gauge invariant metric fields $\gamma_{\mu\nu}$
defined by (\ref{65}) will be given in the relativistic case by
{\be  \dot\gamma_{\mu\nu}=\frac{2}{c^2} u_{(\mu}\dot u_{\nu)}\, ,
\label{71a}\fe}
and in the Newtonian limit by
{\be \dot\gamma_{\mu\nu}=-2 t_{(\mu}\gamma_{\nu)\rho}\dot u^\rho\, ,
\hskip 1 cm \dot\gamma^\mu_\nu=-t_\nu \dot u^\mu\, .\label{71}\fe}

It is useful for many purposes -- and will be indispensible for the
discussion of the Newtonian limit - to introduce an appropriate
fixed measure form, with antisymmetric components 
 $N_{_{ABC}}= N_{_{[ABC]}}$ say, on the rheometric base manifold.
Such a measure will typically be interpretable as representing the 
density of microscopic lattice points in an underlying cristal 
structure. Such a measure will determine a 
corresponding contravariant spacetime current of the kind introduced in  
(\ref{C7}) by a pull back relation of the form
\be 
n^\ast_{\mu\nu\rho}=\,N_{_{ABC}}\,q^{_A}_{\, ,\mu}
q^{_B}_{\, ,\nu}\,q^{_C}_{\, ,\rho}\, ,\label{E12}
\fe
The corresponding
scalar number density $n$ in the rest frame of the medium will
evidently be given by
\be
 n^2=\frac{1}{3!} N_{_{ABC}} N_{_{DEF}}\,\gamma^{_{AD}}
\gamma^{_{BE}}\gamma^{_{CF}} \, .\label{E13}
\fe
It is then obvious that the current density so defined will be automatically 
conserved due to the closure of the 3 form $N_{_{ABC}}$ (being a 3 form on a 
3 manifold):
\be 
\nabla_{\![\lambda}N_{\mu\nu\rho]}=0\, ,\hskip 1 cm
\nabla_{\!\mu} n^\mu=0\, .\label{R43}
\fe

In the simple purely elastic application considered here, the formalism 
defined above will be enough to describe the dynamics of the solid. 
However such a simplification will not be possible in the more general 
applications envisaged for future work, such as transfusive 
exchange~\cite{LSC97} of matter between distinct chemical constituents, 
in which one no longer has conservation of the solid's current.

\bigskip
{\bf 6. Action for a simple perfectly elastic solid}
\medskip

Whereas the internal energy depends only on the density when a simple 
fluid is considered, in the case of a solid it will also be necessary
to allow for the reaction of the internal energy not just to changes of 
volume, but also to changes in shear strain. This is done by allowing  
$\LLambda_{int}$ to depend on $\ggamma^{AB}$, which, because of its time 
dependence can be interpreted as a Cauchy type strain tensor. This kind
of (perfectly elastic) solid model includes the category of simple
(barotropic) perfect fluid models as the special case for which
the dependence is only on the determinant of $\ggamma^{AB}$. For a medium
that is perfectly elastic in this sense, the generic action variation 
will be given in the relativistic case by the formula
\be
 \delta\widetilde\Lambda=\frac{\partial\widetilde\Lambda}{\partial 
\gamma^{_{AB}}}\delta \gamma^{_{AB}}+\frac{\partial\widetilde\Lambda}
{\partial q^{_A}}\delta q^{_A}\, ,\label{E8}
\fe
which characterises the partial derivative components needed for
the specification of the rheometric stress tensor $S_{_{AB}}$ and
its spacetime pullback
{\be S_{\mu\nu}=S_{_{AB}}\,q^{_A}_{\, ,\mu}q^{_B}_{\, ,\nu}
\, ,\hskip 1 cm
S_{_{AB}}=2\frac{\partial\tilde\Lambda}{\partial 
\gamma^{_{AB}}} -\widetilde\Lambda \gamma_{_{AB}}
\, .\label{E9}\fe}
It can be seen from (\ref{C32}) and (\ref{E4}) that the complete stress
energy tensor will be given in terms of this quantity by an expression
of the standard form
{\be \widetilde T{^{\mu\nu}}=\widetilde\rho\, u^\mu u^\nu-S^{\mu\nu}
\, ,\label{E10}\fe}
in which comoving mass-density density $\widetilde\rho$ is given 
by the simple proportionality relation
{\be \widetilde\Lambda= -\widetilde\rho\, c^2\, .\label{E11}\fe}
This is all that is needed for the formulation of the variational
field equations, which in this case are given completely 
just by the Noetherian divergence condition (\ref{C34}).

In typical physical applications the mass-energy function in (\ref{E11}) 
will have a minimum value $\rho_{\rm _{flue}}$ -- corresponding to a 
maximum  value $\Lambda_{\rm _{flu}}$ of the elastic action function 
$\widetilde \Lambda$  -- for any given value of the determinant 
$\vert\gamma\vert$ of the induced metric $\gamma_{_{AB}}$, or 
equivalently for any given value of the number density $n$ as 
specified by (\ref{E13}). This means that the action will
be decomposible in the form
\be 
\widetilde \Lambda= \Lambda_{\rm _{flu}}+
  \Lambda_{\rm _{rig}}\, ,\label{E14}
\fe
in which the -- usually rather small -- remainder term
-- $\Lambda_{\rm _{rig}}$ is a negative indefinite rigidity contribution
contribution without which the medium  would be of purely
fluid type. The -- usually dominant -- perfect fluid contribution
$\widetilde \Lambda_{\rm _{flu}}$ will itself be itself decomposible
in the form 
\be 
\Lambda_{\rm _{flu}}=\Lambda_{\rm_{bal}}+\Lambda_{\rm_{pre}}
\, ,\label{E15}
\fe
in which $\Lambda_{\rm_{pre}}$ is a pressure energy contribution
that in typical applications will also be relatively small 
compared with a dominant ballistic contribution $\Lambda_{\rm_{bal}}$
given by an appropriate choice of the mass parameter $m$ in
the general formula (\ref{C12}) which, in the single constituent case 
considered here, reduce to the trivial form
{\be \Lambda_{\rm_{bal}}= -\rho\, c^2\, , \hskip 1 cm
\rho=m\, n\, ,\label{E16}\fe}
so that the mass density (\ref{E11}) will be expressible in the form
{\be \widetilde\rho=\rho+c^{-2} \Euro \, ,\label{E17}\fe}
with the relatively small internal energy contribution given 
according to (\ref{C11}) by
{\be \Euro=- \Lambda_{\rm_{int}} \, .\label{E18}\fe}
The action decomposition (\ref{E14}) corresponds to an energy density 
decomposition of the form
{\be \Euro=\Euro_{\rm_{pre}}+\Euro_{\rm_{rig}} \, ,\label{E19}\fe}
in which the rigidity contribution $\Euro_{\rm_{rig}}$ will vanish
when $\Euro$ is minimised for a given value of $n$. The non trivial
pressure term 
{\be \Lambda_{\rm_{pre}}=-\Euro_{\rm_{pre}}\, ,\label{E20}\fe}
 in (\ref{E15})
will evidently be a function just of the undifferentiated base space 
coordinates $q^{_A}$ and of the scalar density $n$, which means that, as 
the analogue of (\ref{E8}), its generic variation will be given by
{\be \delta  \Lambda_{\rm _{pre}}=-\frac{\partial  \Euro_{\rm _{pre}}}
{\partial n}\delta n-\frac{\partial  \Euro_{\rm _{pre}}}
{\partial q^{_A}}\delta q^{_A}\, ,\, .\label{E21}\fe}
with
{\be  \delta n=\frac{n}{2}
\gamma_{_{AB}}\delta\gamma^{_{AB}}\, .\label{67}\fe}
It is evident from (\ref{E9}) that the ballistic term will make no 
contribution at all to the strain tensor, while the contribution from the 
pressure energy term will of course be purely isotropic: 
{\be S_{\rm_{bal}}^{\ \mu\nu}=0\, ,\hskip 1 cm S_{\rm_{pre}}^{\ \mu\nu}
=-P_{\rm_{pre}}\gamma^{\mu\nu}\, ,\hskip 1 cm P_{\rm_{pre}}=
 n\frac{\partial  \Euro_{\rm _{pre}}}{\partial n}-\Euro_{\rm_{pre}}\!
\, .\label{E22}\fe}
This means that the purely intrinsic action contribution given, according 
to (\ref{E19}), by
{\be \Lambda_{\rm_{int}}=\Lambda_{\rm_{pre}}+
\Lambda_{\rm_{rig}}\, ,\label{E23}\fe}
will be sufficient by itself to determine the the complete stress tensor 
in (\ref{E10}), which will take the form
{\be S^{\mu\nu}=S_{\rm_{int}}^{\ \mu\nu}=-P_{\rm_{pre}}\gamma^{\mu\nu}
+S_{\rm_{rig}}^{\ \mu\nu}\, .\label{E24}\fe}

It is important to notice that in this simple elastic case the effect of a 
worldline displacement along the worldlines will have no effect
so substitution of $u^\mu$ for $\xi^\mu$ in (\ref{C33}) will merely give an
identity 
{\be u^\mu \widetilde \nabla_{\!\nu}\widetilde T^\nu_{\ \mu}=0 
\label{E24a}\, .\fe}
A complete set of dynamical equations will therefore be provided just by
the orthogonal projection of (\ref{C34}) which will be expressible as
{\be \left(\widetilde \rho \gamma^\mu_{\ \nu}-c^{-2}S^\mu_{\ \nu}\right)
\dot u^\nu=\gamma^\mu_{\ \rho}\gamma^{\sigma\nu}\widetilde\nabla_{\!\nu} 
S^\rho_{\  \sigma}\label{E24b}\, .\fe}

\bigskip
{\bf 7. Newtonian dynamics of a simple perfectly elastic solid.}
\medskip

To obtain the Newtonian limit of the simple relativistic elasticity model
characterised by the stress energy tensor (\ref{E10}) that was set up
in the preceeding section, it is now straightforward to apply the 
procedure described in Section 3, according to which we should 
use a Newtonian action integral 
{\be {\cal I}=\int \Lambda\,
\vurepsilon_{_{1\, 2\, 3\, 0}}\,{\rm d}x^{_1}
\,{\rm d}x^{_2}\,{\rm d}x^{_3}\,{\rm d}x^{_0} \, ,\label{E26}\fe} 
with a Lagrangian of the form (\ref{C27}), which in this simple
elastic case reduces just to
{\be \Lambda=\Lambda_{\rm_{kin}}+\Lambda_{\rm_{pot}} -\Euro
\, ,\label{E27}\fe}
with an internal energy function $\Euro$ of the same form as
in the relativistic case, and with the Newtonian kinetic and
potential energy terms given by expressions of the same form as
for a simple fluid, namely
{\be \Lambda_{\rm_{kin}}= \frac{_1}{^2} \rho\,
v^{2}\, ,\hskip 1 cm \Lambda_{\rm_{pot}}
=-\phi \rho\, ,\label{E28}\fe}
in which the  3-velocity $v^\mu$ is charcterised in terms
of the chosen ether frame frame, $e^\mu$ by
\be v^\mu=u^\mu-e^\mu\, ,\hskip 1 cm v^2=\eta_{\mu\nu}\,u^\mu u^\nu
\, ,\label{E29}\fe
and relevant Newtonian mass density will be given simply by
{\be \rho=m\, n\, ,\label{40}\fe}
for a number density $n$ that, in accordance with
(\ref{E12}),  will be expressible in this limit as
{\be n = n^\mu t_\mu={1\over 3!}
\epsilon^{\mu\nu\rho}n_{\mu\nu\rho}\, ,\hskip 1 cm
n^\mu=n u^\mu={1\over 3!}\vorepsilon^{\mu\nu\rho\sigma}
n^\ast_{\nu\rho\sigma}\, .\label{41}\fe}
The current 3 form $n^\ast_{\nu\rho\sigma}$ itself is obtained from the 
fixed 3- form $N_{_{ABC}}$ on the material base space by the construction
(\ref{E12}), which automatically ensures the satisfaction of the 
non-relativistic conservation law in its usual form
{\be \nabla_{\!\mu} n^\mu=0\, .\label{43}\fe} 
It is to be remarked that, as in the fluid case~\cite{CC02}, the 
gravitational coupling term $\Lambda_{_{\rm pot}}$
will be unaffected by linear (Galilean) gauge transformations
but not by accelerated Milne transformations, while the kinetic term
$\Lambda_{_{\rm kin}}$ is of course even more highly frame
 dependent (not even Galilei invariant).For the purpose of comparison 
with the preceding work, the terms in the Newtonian action can usefully 
be regrouped in the standard form 
{\be \Lambda=\Lambda_{_{\rm mat}}+\Lambda_{_{\rm pot}}\, \, \hskip 1 cm 
\Lambda_{_{\rm mat}}=\Lambda_{_{\rm kin}}-\Euro\, ,\label{48}\fe}

In the perfect fluid case this elastic energy density $\Euro$ will be
given on each worldline (as specified by the values $q^{_A}$) as a 
corresponding function (in the barotropic case~\cite{CC02} everywhere the 
same function) of the number density $n$, which can be seen from (\ref{40})
to be given as a function of the scalar product fields $\gamma^{_{AB}}$ by 
the determinant formula
\be n^2={1\over 3!}N_{_{ABC}}N_{_{DEF}}\gamma^{_{AD}}\gamma^{_{BE}}
\gamma^{_{CF}}\, .\label{50} \fe
The generalisation from a perfect fluid to a perfectly elastic solid is 
made simply by taking  $\Euro$ to be a generic (worldline dependent) 
function of the scalar product fields $\gamma^{_{AB}}$ (i.e. of the 
tensorial projection of $\gamma^{\mu\nu}$ onto $\cal X$) instead of 
being restricted to the depend just on the determinental combination
(\ref{50}) as in the fluid case. This means allowing $\Euro$ to
be affected not just by changes of volume but also by shearing strain,
whose effect is allowed for by supplementing the purely
fluid contribution $\Euro_{\rm pre}$ in (\ref{E18}) by the extra
rigidity term  $\Euro_{\rm rig}$.

Using (\ref{22}) and (\ref{23}) we obtain the formulae
{\be \delta u^\mu=-u^\mu u^\nu\delta t_\nu \label{51}\fe}
and 
{\be \delta n={_1\over ^2} n\gamma_{\mu\nu}\delta \gamma^{\mu\nu} =
n\eta_{\mu\nu}({_1\over^2}\delta\gamma^{\mu\nu}
+u^\mu\gamma^{\nu\rho}\delta t_\rho)\label{52}\fe}
for variations in which the fields $q^{_A}$ are held constant -- meaning 
that the world lines remain fixed -- it can be seen from the defining 
ansatz (\ref{33}) that the kinetic and gravitational potential contributions 
to the stress momentum energy density tensor will be given by expressions 
that are already familiar from experience~\cite{CC02} with the simple 
fluid case, namely
{\be T^{\ \mu}_{\!_{\rm kin}\nu}= n^\mu p_\nu \label{53}\fe}
in which the frame dependent 4- momentum per particle is given, using 
the notation of (\ref{E29}) by
{\be p_\mu=m (v_\mu -{_1\over^2}v^2 t_\mu)\, \label{54}\fe}
and 
{\be T^{\ \mu}_{\!_{\rm pot}\nu}= -\phi\rho^{\mu} t_\nu \, , ~~\rho^{\mu}= 
m n^{\mu}.\label{59}\fe}

The corresponding stress tensor $S^{_{AB}}$ is definable 
in the traditional manner in terms of the effect on the energy per particle 
$\Euro/n$ of an infinitesimal strain variation $\delta\gamma_{_{AB}}$ 
according to a prescription of the form
{\be n\,\delta\Big({\Euro\over n}\Big)={_1\over^2}S^{_{AB}}
\delta \gamma_{_{AB}}\, .\label{72}\fe}
By comparing this with the formula (\ref{67}) for $\delta n$, it can be 
seen that this contravariant stress tensor will be expressible 
independently of the number density as
{\be S^{_{AB}}=2{\partial\Euro\over\partial \gamma_{_{AB}}}+\Euro
\gamma^{_{AB}}\, ,\label{73}\fe}
which means that its covariant (index lowered) version will be given by
{\be S_{_{AB}}=-2{\partial\,\Euro\over\partial \gamma^{_{AB}}}+\Euro
\gamma_{_{AB}}\, .\label{74}\fe}
In the manner described above this base tensor will determine a 
corresponding space stress tensor
{\be S_{\mu\nu}=S_{_{AB}}q^A{_{,\mu}}q^A{_{,\nu}}\, ,\hskip 1 cm
S_{\mu\nu}u^\nu=0\, ,\label{75}\fe}
in terms of which the rate of change of the energy density along the world 
lines will be given by
{\be \dot{\Euro}=S_{\mu\nu}\theta^{\mu\nu}-\Euro\theta\, ,\label{76}\fe}
where $\theta$ is the expansion rate, as given by
{\be \theta=\gamma_{\mu\nu}\theta^{\mu\nu}=\nabla_\mu u^\mu=-\dot n/n
\, .\label{77}\fe}

Using the formula
\be {\partial\gamma^{_{AB}}\over \partial\gamma^{\mu\nu}}=
q^{_A}_{\, ,(\mu}q^{_B}_{\, ,\nu)}\label{78}\fe
obtained from (\ref{47}), it can be now seen from (\ref{72}) that we shall 
have
\be {\partial\Euro\over\partial\gamma^{\mu\nu}}={_1\over^2}(\Euro
\gamma_{_{AB}}-S_{_{AB}})q^{_A}_{\, ,\mu}q^{_B}_{\, ,\nu}
={_1\over^2}(\Euro\gamma_{\mu\nu}-S_{\mu\nu})\, .\label{79}\fe
The relation (\ref{E18}) thus gives
\be{\partial\Lambda_{_{\rm int}}\over\partial\gamma^{\mu\nu}} 
={_1\over^2}(S_{\mu\nu}-\Euro\gamma_{\mu\nu})\, ,\label {80}\fe
so that by the general formula (\ref{33})(since no dependence on $e^\mu$ or 
$t_\mu$ is involved in this case) the internal contribution to the stress 
momentum energy tensor will be given by
\be T^{\ \mu}_{\!_{\rm int}\nu}=-\Euro\delta^\mu_\nu
+\gamma^{\mu\rho}(\Euro\gamma_{\rho\nu}-S_{\rho\nu})\, ,\label{81}\fe
which simplifies, using (\ref{E5}) and the (orthogonal) index raising
operation exemplified by (\ref{68}) to
\be T^{\ \mu}_{\!_{\rm int}\nu}=-\Euro u^\mu t_\nu
-S^\mu_{\ \nu}\, .\label{82}\fe
Thus, combining this gauge independent internal contribution with the
ether frame dependent kinetic contribution (\ref{53}), we end up with the 
complete material stress energy tensor
\be T^{\ \mu}_{\!_{\rm mat}\nu}=T^{\ \mu}_{\!_{\rm kin}\nu}
+T^{\ \mu}_{\!_{\rm int}\nu}\, ,\label{83}\fe
in the form
\be T^{\ \mu}_{\!_{\rm mat}\nu} = n^\mu m_\nu
-S^\mu_{\ \nu}\, .\label{84}\fe
in which the relevant (gauge dependent) momentum per particle covector 
$m_\mu$ is given by
\be m_\nu=p_\nu-\big({\Euro/ n}\big)t_\nu=m\left(v_\nu-\big(
{_1\over^2}v^2+{\Euro/\rho}\big)t_\nu\right) \, .\label{85}\fe
Since it is not orthogonal to the flow this momentum 1-form is not 
completely determined just by the corresponding -- purely kinematic --
contravariant momentum covector, namely
\be m^\mu=\gamma^{\mu\nu}m_\nu=m v^\mu\, .\label{86}\fe

In the special case of the perfect fluid limit we shall simply have
$S^{\mu\nu}=-P\gamma^{\mu\nu}$, where $P$ is the ordinary scalar pressure.
It is important to note, in this case, that the momentum covector $m_\nu$ 
introduced here will not be quite the same as the material momentum covector 
$\mu_\nu$ that was introduced in the preceding work~\cite{CC02}, 
since $m_\nu$ is defined in terms of the integrated internal energy per 
particle, namely $\Euro/n$, whereas $\mu_\nu$ was defined in terms of 
the differential internal energy per particle, namely the chemical 
potential, $\chi=(\Euro+P)/n$. This means that, in the (barotropic) perfect 
fluid limit, the relevant material momentum 1-form will be given by
$\mu_\mu=m_\mu-(P/n)t_\mu$.

The complete stress energy tensor, including allowance for the gravitational
background, can now be obtained as
\be T^{\mu}_{\ \nu}=T^{\ \mu}_{\!_{\rm mat}\nu}
+T^{\ \mu}_{\!_{\rm pot}\nu}\, ,\label{87}\fe
which by (\ref{59}) finally gives
\be T^{\mu}_{\ \nu}=n^\mu(m_\nu-m\phi t_\nu)-S^\mu_{\ \nu}\, .
\label{88}\fe
for the tensor whose divergence will provide the required dynamical
equations according to the Noether relation (\ref{36}).

For a compound system (as exemplified by the muticonstituent fluid models 
studied in the preceding work~\cite{CC03}) the 4-independent components of 
the Noether relation (\ref{36}) would not by themselves be sufficient to 
fully determine the dynamical evolution. However for a simple 
medium such as we are considering here, for which the dynamics are completely 
describable just in terms of the motion of the world lines, as given by the 
evolution of the 3 independent scalar fields $q^{_A}$, the evolution is 
actually overdetermined by the 4 components of (\ref{36}), whose 
contraction with 4-velocity,  namely
\be u^\nu f_\nu= 0\, , \hskip 1 cm f_\nu=
(\nabla_{\!\mu} T^{\mu}_{\ \nu}+\rho\nabla_{\!\nu}\phi) 
=0\, ,\label{89}\fe
merely gives the kinematic identity (\ref{75}), which must be satisfied as 
a mathematical necessity, even under circumstances more general than those 
considered here, in which the space components of the force density $f_\mu$ 
acting on the medium might not be zero.The underlying reason for this 
identity is that whereas the dynamical equations were needed for the 
derivation of the Noetherian condition (\ref{34}) for an arbitrary 
displacement vector field $\xi^\nu$, it  would evidently hold as a mere 
identity for a displacement along the flow lines, i.e. for 
$\xi^\nu\propto u^\nu$, in the simple elastic case. It is to be remarked 
that the logic could be reversed as, was done in the cited work on the 
relativistic case, which started~\cite{CQ72} by postulating the analogue 
of (\ref{89}) as a condition needed for consistency, and then went 
on~\cite{C73} to derive the action formulation as a consequence.

The upshot is that the complete system of dynamical evolution equations
for a simple elastic solid model will be given just by the 3 independent
components of the space projection of (\ref{36}) namely
\be \gamma^{\mu\nu}(\nabla_{\!\rho} T^{\rho}_{\ \nu}+\rho\nabla_{\!\nu}
\phi) =0 \, .\label{90}\fe
Using the conservation law (\ref{43}), it can be seen from the formula 
(\ref{88}) that in terms of the frame dependent gravitational field vector,
\be g^\mu=-\gamma^{\mu\nu}\nabla_{\!\nu}\phi\label{91}\fe
this system will be expressible in the form
\be \rho(\dot u^\mu-g^\mu)=\nabla_{\!\nu}S^{\mu\nu}\, ,\label{92}\fe
in which it is to be observed that each side (though not the separate terms 
on the left) is satisfactorily invariant, not just under linear Galilean 
tranformations, but even under arbitrarily accelerated Milne 
transformations.

\vfill\eject

\bigskip
{\bf 8. Derivation of the characteristic equation.}
\medskip

As in the relativistic case~\cite{C72}, let us now seek the conditions 
governing the a covector $\lambda_\mu$ say that is normal to a 
characteristic hypersurface across which the relevant field quantities 
$n$, $u^\mu$, and $S^{\mu\nu}$ have discontinuous gradients, using the 
standard method of Hadarmard, which exploits the condition that the 
discontinuity of the gradient of a continuous scalar field must be 
proportional to the normal covector $\lambda_{\mu}$, so that in particular 
for the density we shall have
\be[\nabla_{\!\mu}n ]=\hat n \lambda_\mu\, ,\label{93}\fe
for some corresponding scalar discontinuity amplitude $\hat n$. 
The associated unit propagation vector $\nu^\mu$ characterised by
{\be \gamma^{\mu\nu}\nu_\nu \nu_\mu=1\, ,
\hskip 1 cm \nu_\mu u^\mu=0\, ,\label{96a}\fe}
and the propagation velocity, $\upsilon$ say, relative to the local rest frame,
of the discontinuity 
are specifiable by taking $\lambda_\mu$ to have the standard normalisation 
so that it takes the form
{\be \lambda_\mu=\nu_\mu+\upsilon \, c^{-2} u_\mu\, ,\label{96b}\fe}
in the relativistic case, and hence
{\be \lambda_\mu=\nu_\mu-\upsilon t_\mu\,  ,\label{96}\fe}
in the Newtonian limit.
In terms of the same discontinuity covector as in (\ref{93}) the
discontinuity of the gradient of $u^\mu$ will be given by an expression of 
analogous form,
{\be[\nabla_{\!\mu}u^\nu]=\hat u^\nu\lambda_\mu\, ,\label{94} \fe}
in terms of a corresponding vectorial discontinuity amplitude $\hat u^\nu$, 
while similarly for the stress tensor we shall have
{\be[\nabla_{\!\mu}S^{\nu\rho}]=\hat S^{\rho\nu}\lambda_\mu
\, .\label{95}\fe}

Since the evolution of $n$ and $S^{\mu\nu}$ is kinematically
determined by that of the flow lines, the corresponding gradient 
discontinuity amplitudes $\hat n$ and $\hat S^{\nu\rho}$ will be 
determined by the velocity gradient discontinuity amplitude $\hat u$. In 
the case of the number density $n$ it can be seen from (\ref{77}) that we 
shall have
{\be u^\mu\nabla_{\!\mu} n =\dot n =-\rho\theta=- n\nabla_{\!\mu}
 u^\mu\, \label{97}\fe}
so by taking the discontinuity we obtain
{\be u^\mu\lambda_\mu \hat n =- n\lambda_\mu  \hat u^\mu
\, .\label{98}\fe}
The normalisation conditions $u^\mu u_\mu=-c^2$ in the relativistic case and
$u^\mu t_\mu=1$ in the Newtonian case imply corresponding restrictions
$\hat u^\mu u_\mu=0$, $\hat u^\mu t_\mu=0$ respectively, with the implication
that for $\hat u^\mu$, as for $\nu_\mu$ we can 
unambigously and reversibly raise and lower the indices by contraction 
with $\gamma^{\mu\nu}$ and $\gamma_{\mu\nu}$. It follows that
(\ref{98}) will reduce to the simple form
{\be \upsilon\,\hat n=n\,\nu_\mu\hat u^\mu\, .\label{99}\fe}
To write the corresponding relation for $S^{\mu\nu}$ we need the relevant 
elasticity tensor, which is defined in such a way as to have the symmetry 
properties
\be  E^{_{ABCD}}=E^{_{CDAB}}= E^{_{(AB)(CD)}}\, ,\label{100}\fe
by the ansatz
\be E^{_{ABCD}}=4n{\partial^2\over\partial\gamma_{_{AB}}
\partial\gamma_
{_{CD}}}\Big({\Euro\over n}\Big)=2{\partial S^{_{AB}} \over 
\partial \gamma_{_{CD}}}+S^{_{AB}}\gamma^{_{CD}}\, .\label{101}\fe
In terms of this highly symmetric elasticity tensor, the less highly 
symmetric Hadamard elasticity tensor that will be needed below is 
specifiable as
\be  A^{_{ABCD}}=A^{_{CDAB}}= E^{_{ABCD}}+\gamma^{_{AC}}S^{_{BD}}
\, .\label{102}\fe

It follows from (\ref{101})  that the time derivative of the stress tensor 
in the material base space will be given in terms of that of the strain 
tensor by
\be \dot S^{_{AB}}={_1\over ^2}(E^{_{ABCD}}-S^{_{AB}}\gamma^{_{CD}})
\dot\gamma_{_{CD}}\, .\label{103}\fe
For the purpose of evaluating the time derivatives of the corresponding
space time tensors, contravariant base tensors are less convenient than
the corresponding covariant tensors, whose time derivative can be seen,
by (\ref{E0bis}), to pullback directly onto the corresponding Lie derivative
in the manner illustrated in the case of the stress as
\be q^{_A}_{\, ,\mu}q^{_B}_{\, ,\nu}\dot S_{_{AB}}=\vec u\Libra S_{\mu\nu}
=u^\rho\nabla_{\!\rho}S_{\mu\nu}+2S_{\rho(\mu}\nabla_{\nu)}u^\rho
\, .\label{104}\fe
We therefore need the formula obtained by swapping covariant with
contravariant indices in (\ref{103}) which gives
 \be \dot S_{_{AB}}=-{_1\over ^2}(E_{_{ABCD}}-S_{_{AB}}\gamma_{_{CD}}
+4 S_{_{C(A}}\gamma_{_{B)D}})\dot\gamma^{_{CD}}\, ,\label{105}\fe 
from which, by (\ref{64}) we obtain
\be q^{_A}_{\, ,\mu}q^{_B}_{\, ,\nu}\dot S_{_{AB}}=
(E_{\mu\nu\rho\sigma}-S_{\mu\nu}\gamma_{\rho\sigma} +4 S_{\rho(\mu}
\gamma_{\nu)\sigma})\theta^{\rho\sigma}\, .\label{106}\fe
Combining this with (\ref{104}) and using the definition (\ref{65})
of the expansion rate tensor $\theta^{\mu\nu}$ we obtain an evolution
equation for the stress tensor in the form
\be u^\rho\nabla_{\!\rho}S_{\mu\nu}=-2S_{\rho(\mu}\nabla_{\nu)}u^\rho
+(E_{\mu\nu\rho}{^\sigma} -S_{\mu\nu}\gamma_{\rho}^{\ \sigma}
+2 S_{\rho(\mu}\gamma_{\nu)}^{\ \sigma}+ 2 S^\sigma_{\ (\mu}
\gamma_{\nu)\rho})\nabla_{\!\sigma}u^\rho \, .\label{107}\fe
Taking the discontinuity of the gradients in this relation we obtain, as 
the analogue of (\ref{98}),
\be u^\rho\lambda_{\rho}\hat S_{\mu\nu}=-2S_{\rho(\mu}\lambda_{\nu)}
\hat u^\rho+(E_{\mu\nu\rho}{^\sigma} -S_{\mu\nu}\gamma_{\rho}^{\ \sigma}
+2 S_{\rho(\mu}\gamma_{\nu)}^{\ \sigma}+2 S^\sigma_{\ (\mu}
\gamma_{\nu)\rho})\lambda_{\!\sigma}\hat u^\rho\, .\label{108}\fe
After projecting out the time component by contraction with
$\gamma^{\lambda\mu}$, this leaves, as the analogue of (\ref{99}), an 
expression giving the stress gradient discontinuity amplitude as a function 
of the velocity gradient discontinuity in the form
\be \upsilon\,\gamma^{\mu\rho}\gamma^{\nu\sigma}
\hat S_{\rho\sigma}=-(E^{\mu\nu\rho}{_\sigma}
-S^{\mu\nu}\gamma^\rho_{\ \sigma}+2S_\sigma^{\ (\mu}
\gamma^{\nu)\rho})\nu^\sigma\hat u_\rho, .\label{109}\fe

We now have all that is needed for processing the gradient discontinuity 
relation provided by the equations of motion, from which one obtains 
the dynamical equation
\be \rho u^\rho\lambda_\rho\,y^\mu_{\ \nu}\hat u^\nu=
\lambda_\nu\gamma^{\mu\rho}
\gamma^{\nu\sigma}\hat S_{\rho\sigma}\, ,\label{110}\fe
in terms of a tensor  $y^\mu_{\ \nu}$ that will be given
in the relativistic case (\ref{E24b}) by the relation
{\be \rho c^2 y^\mu_{\ \nu}=(\rho c^2+\Euro)\gamma^\mu_{\ \nu}
-S^\mu_{\ \nu}\, ,\label{110a}\fe}
but the reduces in the Newtonian limit (\ref{92}) simply  to
$y^\mu_{\ \nu}=\gamma^\mu_{\ \nu}$.

By substitution of the kinematic formula (\ref{109}) into the dynamical 
condition (\ref{110}) we finally obtain the required characteristic equation 
in the form
\be (\upsilon^2\rho\, y^{\mu\nu}-Q^{\mu\nu})\hat u_\nu=0\, .\label{112}\fe
This  is an effectively 3-dimensional eigenvector equation with $\upsilon^2$, 
the square of the relative propagation speed, as eigenvalue, for which the 
eigenvector is the covariant velocity gradient discontinuity amplitude 
$\hat u_\mu$ as characterised by the orthogonality condition
\be  u^\mu \hat u_\mu=0\, .\label{113}\fe 
In terms of the Hadamard elasticity tensor specified according to 
(\ref{102}), the characteristc matrix $Q^{\mu\nu}$ can be seen to
be expressible as a function of the propagation direction, 
as indicated by the spacelike unit covector $\nu_\mu$ by the formula
\be Q^{\mu\nu}=A^{\mu\rho\nu\sigma}\nu_\rho\nu_\sigma\, .\label{114}\fe
The simplest application of this formula is of course to the case of a
medium that is intrinsically isotropic (as will typically be the case in 
macroscopic applications involving averaging over a large number of randomly 
oriented mesoscopic crystals) and that is in an undeformed, though perhaps 
highly compressed state, with energy density $\Euro_0$ say. In such an 
undeformed state the stress tensor $S^{\mu\nu}$ will reduce to an undeformed 
value, $S_0^{\,\mu\nu}$ say, that will be characterised just by a pressure 
scalar $P_0$, in terms of which it will take the form
\be S_0^{\,\mu\nu}=-P_0\gamma^{\mu\nu}\, .\label{115}\fe
It follows that the tensor $y^{\mu\nu}$ in (\ref{112}) will reduce to 
an undeformed value $y_0^{\,\mu\nu}$ given in the relativistic case by
{\be y_0^{\,\mu\nu}=\left(1+\frac{\Euro_0+P_0}{\rho\, c^2}\right)
\gamma^{\mu\nu}\, .\label{115a}\fe}
As discussed in the cited work~\cite{CQ72} on the relativistic 
case, for such a state the elasticity tensor $E^{\mu\nu\rho\sigma}$ will 
reduce to a corresponding isotropic value $E_0^{\,\mu\nu\rho\sigma}$ that 
is expressible in the well known form
\be E_0^{\,\mu\nu\rho\sigma}=(\beta_0-{_1\over^3}P_0)\gamma^{\mu\nu}
\gamma^{\rho\sigma}+2(\mu_0+P_0)(\gamma^{\mu(\rho}\gamma^{\sigma)\nu}
-{_1\over^3}\gamma^{\mu\nu}\gamma^{\rho\sigma})\, ,\label{116}\fe
in which the coefficients $\beta_0$ and $\mu_0$ are respectively interpretable 
as the bulk modulus and the modulus of rigidity. In the particular case of a 
perfect fluid the rigidity will vanish, $\mu=0$, and the bulk modulus will 
be given by the derivative of the pressure with respect to fractional volume 
change, $\beta=n {\rm d}P/{\rm d}n$. According to (\ref{102}) the Hadamard 
elasticity tensor $A^{\mu\nu\rho\sigma}$ reduce to a corresponding isotropic 
limit value given by
\be A_0^{\,\mu\nu\rho\sigma}=\beta_0\gamma^{\mu\nu}\gamma^{\rho\sigma}
+2 P_0\gamma^{\mu[\sigma}\gamma^{\nu]\rho}+2\mu_0(\gamma^{\mu(\rho}
\gamma^{\sigma)\nu}-{_1\over^3}\gamma^{\mu\nu}
\gamma^{\rho\sigma})\, ,\label{117}\fe
which is such that the antisymmetric pressure term will cancel out in the 
formula for the characteristic matrix $Q^{\mu\nu}$, leaving an expression 
of the same form $Q_0^{\,\mu\nu}$ as is familiar in the low pressure limit, 
namely
\be Q_0^{\,\mu\rho}=(\beta_0+{_1\over^3}\mu_0)\nu^\mu \nu^\rho+
\mu_0\gamma^{\mu\rho}\, .\label{118}\fe

\bigskip
{\bf 9. Faraday - Ampere magnetodynamics.}
\medskip

So long as it acts merely as a given prescribed background, an 
electromagnetic field can be treated within a Newtonian framework in a 
manner that is satisfactorily Galilei and even Milne invariant~\cite{CG88}.
However when it is necessary to treat it in its own right as an active
dynamical field governed by an unrestricted electric current source $j^\mu$
then it is necessary to sacrifice Galilean (and hence a fortiori Milne)
invariance, which was brutally violated by the introduction of a physically 
preferred ether in Maxwell's original formulation, but more elegantly 
replaced by Lorentz invariance in Einstein's special relativistic treatment.

A satisfactorily Galilean and even Milne invariant formulation in terms of 
an antisymmetric field 2-form  with components $F_{\mu\nu}=-F_{\nu\mu}$ can 
however be set up as a self consistent approximation in cases for which 
only a subset of three ``magnetic'' degrees of freedom are dynamically 
independent, while the other three ``electric'' components (out of the six 
contained in $F_{\mu\nu}$) and all the components of the current are treated 
merely as passively derived fields in the manner that will be described 
immediately below in this section.

In the following section it will be shown how such an ether gauge invariant 
Faraday-Ampere type  model can be coupled, in a variational formulation, to 
a simple perfect barotropic fluid or elastic solid model, of the kind 
described in the previous section, in the special case  of  ``perfect 
conductivity'', meaning the case for which the  field is ``purely magnetic'' 
in the sense that, with respect to the local rest frame specified by 
the velocity 4 vector $u^\mu$ of the fluid, the relevant electric components 
are zero. We thus obtain a 4-dimensionally covariant formulation of the 
non - relativistic version of what is known in the fluid 
case~\cite{BekOron00} as a perfect magnetohydrodynamics.

In the generic case, the electric and magnetic fields $E_\mu$ and 
$B_{\mu\nu}$ can be defined, with respect to an ether vector $e^\mu$,
by the decomposition
\be F_{\mu\nu}=B_{\mu\nu}+2E_{[\mu}t_{\nu]}\, ,\label{120}\fe
subject to the conditions 
\be E_\mu e^\mu=0\, ,\hskip 1 cm B_{\mu\nu}e^\nu=0\, ,\label{121}\fe
which are equivalent to the specification
\be E_\mu=F_{\mu\nu}e^\nu\, ,\label{122}\fe
where, in the Newtonian case we are concerned with here, $t_\mu$ is the 
preferred time covector introduced in (\ref{1}) (while in the relativistic 
case it would be given in terms of the spacetime metric by 
$t_\mu=-g_{\mu\nu}u^\mu$). Under the action of an ether gauge 
transformation of the form (\ref{10}), as generated by a spacelike 
boost vector field $b^\mu$, these fields will acquire new values given by
\be \breve E_\mu=E_\mu+ (B_{\mu\nu}-t_\mu E_\nu)b^\nu\, , \hskip 1cm
\breve B_{\mu\nu}=B_{\mu\nu}+2t_{[\mu} B_{\nu]\rho} b^\rho\, . 
\label{123}\fe
but the gauge dependence of the corresponding contravectorial quantities
\be E^\mu=\gamma^{\mu\nu}E_\nu\, ,\hskip 1 cm B^\mu
={_1\over^2}\epsilon^{\mu\nu\rho}B_{\nu\rho}\, ,\label{124}\fe
will be simpler, so much so that the vector $B^\mu$ will actually be 
physically well defined in the sense of being independent of the  choice 
of ether frame, since we shall have
\be \breve E^\mu=E^\mu+\gamma^{\mu\nu}B_{\nu\rho}b^\rho\, ,
\hskip 1 cm \breve B^\mu=B^\mu\label{125}\, .\fe
The kinematic field 2-form closure condition
\be \nabla_{\![\mu}F_{\nu\rho]}=0\, ,\label{126}\fe
corresponds two of the 4-Maxwell equations, which are expressible in our 
4-dimensionally covariant notation scheme as
\be \nabla_\mu B^\mu=0\, ,\hskip 1 cm \epsilon^{\mu\nu\rho}
\nabla_{\!\nu}E_\rho=-e^\nu\nabla_{\!\nu}B^\mu\, ,\label{127}\fe
of which the second is interpretable as Faraday's law of magnetic
induction.

The trouble, in a Newtonian framework, is with the other two Maxwell 
equations, which specify the way an arbitrary source current 4-vector 
$j^\mu$ governs the dynamic evolution of the field. In a relativistic 
theory this is done by setting $\nabla_{\!\nu}F^{\mu\nu}=4\pi j^\mu$, 
where $F^{\mu\nu}$ is obtained from $F_{\mu\nu}$ by contraction with the 
non-degenerate contravariant spacetime metric $g^{\mu\nu}$. However the 
analogous Newtonian procedure of contraction with $\gamma^{\mu\nu}$ 
will, due to the degeneracy of the latter, give a result that is 
overdetermined, having a form that is expressible in terms
of the rationalised magnetic field
\be H^{\mu\nu}={1\over 4\pi}\gamma^{\mu\rho}\gamma^{\nu\sigma}F_{\rho\sigma}
={1\over 4\pi}\gamma^{\mu\rho}\gamma^{\nu\sigma}B_{\rho\sigma}
\, ,\label{128}\fe
which is ether gauge independent, and purely spacelike,
\be \breve H^{\mu\nu}= H^{\mu\nu}\,~~ H^{\mu\nu}t_\nu=0\, ,\label{129}\fe
as the Ampere type equation
\be \nabla_{\!\nu}H^{\mu\nu}=j^\mu\, .\label{130}\fe
It is evident from (\ref{129}) that this can be satisfied only if the 
current is restricted  to be similarly spacelike, in the sense of 
satisfying the consistency condition
\be j^\mu t_\mu=0\, .\label{131}\fe
The associated electromagnetic action density $\Lambda_{_{\rm F}}$, as 
similarly obtained from the usual relativistic action density 
$F^{\mu\nu}F_{\nu\mu}/16\pi$ by substituting $\gamma^{\mu\nu}$ for
$g^{\mu\nu}$, will have the form
\be \Lambda_{_{\rm F}}=-\Euro_{_{\rm F}} \, ,\label{132}\fe
where $\Euro_{_{\rm F}}$ is the ether gauge independent magnetic energy 
density as given by
\be \Euro_{_{\rm F}}={B^2\over 8\pi}\, ,\hskip 1 cm
B^2=\eta_{\mu\nu}B^\mu B^\nu={_1\over^2} \gamma^{\mu\rho}
\gamma^{\nu\sigma}F_{\mu\nu}F_{\rho\sigma}\, ,\label{133}\fe
from which, by considering the effect of varying the 
degenerate metric $\gamma^{\mu\nu}$ for fixed $F_{\mu\nu}$,
the corresponding stress energy tensor is obtainable according to
the ansatz (\ref{33}) in the form
\be T^\mu_{\!_{\rm F}\,\nu}=H^{\mu\rho}F_{\nu\rho}-\Euro_{_{\rm F}}
\delta^\mu_\nu\, .\label{134}\fe 
The force density acting on the field (the opposite of the 
Faraday - Lorentz type electromagnetic  reaction on the relevant medium) 
will therefore be given by the expression
\be f_{_{\rm F}\nu}=\nabla_{\!\mu} T^\mu_{\!_{\rm F}\,\nu}=j^\mu
F_{\mu\nu}\, .\label{135}\fe

This necessary restriction (\ref{131}) is interpretable as meaning that 
there can be no net electric charge density, a condition which replaces 
the traditional Coulomb equation that would be expressed in our covariant 
notation scheme as $\nabla_{\!\nu}E^\nu=4\pi j^\mu t_\mu$, but which can 
be seen from (\ref{125}) to be incompatible with Galilean invariance unless 
the magnetic part of the field is absent, and which even then will 
be
incompatible with (\ref{123}) except in the pure vacuum case for which
there is no source current $j^\mu$ at all. 

To set up an ether frame invariant model for use as an exactly self
consistent approximation in a Newtonian framework we are thus faced with
a choice between two generically incompatible alternatives. One
possibility (which is likely to be most realistic when insulating material
is involved) is to use a scheme based on the Coulomb equation, which
entails abandoning the Ampere equation and simply restricting the magnetic
part of the field to be zero. The other possibility (more likely to be
realistic for dealing with good conductors) which is the option chosen
for the present work, is to use a scheme based on the Ampere equation
(\ref{130}) in conjunction with the force law (\ref{135}),  which entails
abandoning the Coulomb equation and restricting the charge density to be
zero in accordance with (\ref{129}). This effectively demotes the current
from the status of an independent dynamical variable to that of a derived
quantity, and entails a concomitant loss of independence of the electric
part of the field, which instead of the Coulomb equation, is required, in
the most commonly used kind of model, to obey an Ohm type equation, of
which the simplest version is covariantly expressible, in terms of the
4-velocity vector $u^\mu$ of the relevant supporting -- fluid or solid --
medium, as $\gamma^{\mu\nu}F_{\mu\nu} u^\nu=\kappa j^\mu$,
where $\kappa$ is a resistivity scalar that, in the case of a non-isotropic
solid, might need to be replaced by a tensor. For positive
resistivity, $\kappa>0$, such an Ohm ansatz can be applied in the case of a
composite medium involving entropy density as an independent degree of 
freedom, but its substitution in the force law (\ref{135}) shows that
it will entail a generically positive rate of energy transfer to the medium 
that will be given as a quadratic function of the (purely spacelike) current 
(\ref{130}) by an expression of the form $u^\nu f_{_{\rm F}\nu}=\kappa j^2$,
where $j^2=\eta_{\mu\nu}j^\mu j^\nu$. It is evident however that this will not
in general be compatible with the identity (\ref{89}) that must be satisfied
for a single constituent medium of the kind to which the present study is
restricted. To obtain a self consistent model involving just a simple solid or
(barotropic) fluid, we need to restrict ourselves to the non-dissipative
perfectly conducting case case for which the resistivity vanishes, $\kappa=0$.

\bigskip
{\bf 10. Perfect magneto - elastic dynamics.}
\medskip

It is evident from the foregoing considerations that the perfect conductivity
condition needed to characterise a medium of the simple non dissipative kind
considered here reduces to the perfect conductivity condition that is 
expressible covariantly as the condition
\be F_{\mu\nu}u^\mu=0\, ,\label{137}\fe
which is interpretable as meaning that with respect to the local rest frame 
specified by the material 4-velocity $u^\mu$ the field is of a purely 
magnetic character. This condition is not just mathematically convenient 
but also justifiable -- due to the relatively small mass of the electrons 
that are typically the main charge carriers -- as a remarkably good
approximation in many terrestrial applications and in a very wide range of 
astrophysical contexts, of which the most extensively studied so far have
been those for which the the relevant material medium is a simple
perfect fluid, in which case the ensuing theory is what is known
as perfect magnetohydrodynamics. As has been shown by the work of Jacob 
Bekenstein with Eleizer and Asaf Oron~\cite{BekOron00,BekOron78}, the 
elegant mathematical properties of this kind of magnetohydrodynamic model 
are easier to analyse in a fully relativistic framework. Part of
the motivation for the 4-dimensionally covariant approach developed here is 
to facilitate the extra work needed~\cite{BekOron00} for the treatment
of the Newtonian limit. It is to be noted that the variational formulation 
developed below differs, in the fluid limit, from the one developed by 
Bekenstein and Oron~\cite{BekOron00} who worked with Clebsch type potentials 
of the kind introduced in a relativistic context by Schutz~\cite{Schutz}. 
The use of such Clebsch type variables is just one of several possibilities 
that may be convenient for various purposes in a purely fluid context, but 
like most of the other alternatives it has the disadvantage of being 
unsuitable for generalisation to solids. For the purpose of setting up a
variational formulation for the treatment of an elastic solid medium it 
has long been clear~\cite{C73} that the only practical option is to work 
in terms of world line displacements as characterised by comoving coordinate 
variable of the kind denoted here by $q^{_A}$ .

As a consequence of the closure property (\ref{126}),  it follows that the 
2 - form $F_{\mu\nu}$ will be ``frozen in'' in the sense of having 
vanishing Lie derivative with respect to the flow:
\be \vec u \Libra\, F_{\mu\nu}\equiv u^\rho\nabla_{\!\rho}
F_{\mu\nu}+2 F_{\rho[\nu}\nabla_{\!\mu]} u^\rho=0\, .\label{138}\fe
 since an antisymmetric matrix cannot have even rank, the orthogonality 
condition (\ref{137}) implies that, as well as $u^\mu$, 
the field $F_{\mu\nu}$ possesses another independent null eigenvector,
which can be taken to be $B^\mu$ as given by (\ref{124}). A well known
consequence is that the Maxwellian 2-form $F_{\mu\nu}$ will be conserved 
by Lie transport along any vector that is a linear combination of the form
 $\xi^\mu = \alpha_1 B^\mu+\alpha_2 u^\mu$ where $\alpha_1$ and $\alpha_2$ 
are any scalar fields, and it can also be seen~\cite{C89} that the 2-surface 
elements spanned by such vectors will mesh together to form a congruence
of well defined flux 2-surfaces.

In the same way as remarked above about the stress tensor $S_{\mu\nu}$, 
the world line orthogonality property (\ref{137}) is interpretable as 
meaning that $F_{\mu\nu}$ naturally determines and is determined by a 
corresponding antisymmetric material base tensor with components $F_{_{AB}}$ 
such that
\be F_{\mu\nu}=F_{_{AB}}q^{_A}_{\, ,\mu} q^{_B}_{\, ,\nu}
\, ,\label{139}\fe
while the Lie transport condition (\ref{138}) is interpretable as
meaning that this induced field will be time independent, 
\be \dot F_{_{AB}}=0\, \label{140}\fe
so that the covariant components $F_{_{AB}}$ will be those of a fixed 
2-form field on the 3 dimensional base manifold $\cal X$. Morever it can 
be seen that as a consequence of the space-time closure property
(\ref{126}) the base space 2-form field will have a corresponding closure 
property,
\be F_{[_{AB} ,_C]}=0\, ,\label{141}\fe
which means that it will locally be expressible as the exterior derivative,
\be F_{_{AB}}=2A_{[_B ,_A]}\, ,\label{142}\fe
of a ``frozen in'' 1-form field with components $A_{_A}$ of the kind 
introduced in (\ref{62}), whose spacetime pull back (\ref{68}) thereby
provides an expression of the familiar form
\be F_{\mu\nu}=2\nabla_{\![\mu}A_{\nu]}\, .\label{142a}\fe
It is to be remarked that this natural material gauge is not uniquely defined, 
since there is still some liberty in the choince of $A_{_B}$. 
As a result of the degeneracy property (\ref{137}), there exists a 
current $\eta^\mu$ defined by
\be \eta^\mu = {_1\over^2}\varepsilon^{\mu\nu\rho\sigma} A_\nu 
F_{\rho\sigma}, \fe 
that is conserved in the sense of satisfying
\be \nabla_\mu \eta^\mu =\varepsilon^{\mu\nu\rho\sigma} F_{\mu\nu} 
F_{\rho\sigma} = 0, \fe 
and that has a time component
\be \eta^\mu t_\mu = -A_\mu B^\mu \fe 
which is proportional to the well known magnetic helicity 
scalar \cite{Woltjer58}. 

As a consequence of this ``frozen in'' behaviour, it can be seen
that allowance for the effect of the magnetic field can be included
directly within the perfect elasticity formalism developed in Section 5 
simply by taking the energy density to have the form
\be \Euro=\Euro_0+\Euro_{_{\rm F}}\label{143}\fe
in which $\Euro$ is the purely material part -- depending just on the base 
coordinates $q^{_A}$ and the induced metric components $\gamma^{_{AB}}$ --
 that would remain when the field components $F_{_{AB}}$ components are set 
to zero, while the other part $\Euro_{_{\rm F}}$ is specified as a function 
not just of $q^{_A}$ and $\gamma^{_{AB}}$ but also of the field components 
$F_{_{AB}}$ whose status will be that of initial data that -- subject to 
the closure condition (\ref{141}) are freely specifiable, but that once 
chosen will evolve as fixed functions of the base coordinates $q^{_A}$. 
For a fixed base value of the base coordinates $q^{_A}$ the most general 
variation of $\Euro$ will determine not just a corresponding (symmetric) 
stress tensor $S^{_{AB}}$ but also a corresponding (antisymmetric) magnetic
field tensor by the prescription
\be \delta \Euro = {_1\over^2}(S^{_{AB}}-\Euro\gamma^{_{AB}})\delta
\gamma_{_{AB}}+{_1\over ^2}H^{_{AB}}\delta F_{_{AB}}\, ,\label{144}\fe
which decomposes with
\be S^{_{AB}}= S_0^{\,_{AB}}+S_{_{\rm F}}^{\,_{AB}}\, ,\label{145}\fe
as the sum of parts given by
 \be \delta \Euro_0 = {_1\over^2}(S_0^{\,_{AB}}-\Euro_0\gamma_{_{AB}}
)\delta \gamma_{_{AB}}\, ,\label{146}\fe
and
\be \delta \Euro_{_{\rm F}} = {_1\over^2}(S_{_{\rm F}}^{\,_{AB}}-
\Euro_{_{\rm F}}\gamma^{_{AB}})\delta\gamma_{_{AB}}\,
+{_1\over ^2}H^{_{AB}}\delta F_{_{AB}}\, .\label{147}\fe

In the general case of a polarised medium the functional form of
$\Euro_{_{\rm F}}$ might be rather elaborate, but in the simple case of 
an unpolarised medium it is simply to be identified with the ordinary 
vacuum magnetic energy density as given by (\ref{133}) which is 
translatable into terms  material base fields as
\be \Euro_{_{\rm F}}={_1\over ^4}H^{_{AB}}F_{_{AB}}\, ,\label{148}\fe
where, consistently with (\ref{147}) the magnetic field tensor
$H^{_{AB}}$ is given in this particular case simply by
\be H^{_{AB}}={1\over 4\pi}\gamma^{_{AC}}\gamma^{_{BD}}F_{_{CD}}
\, .\label{149}\fe
while the corresponding magnetic stress contribution will be obtainable 
from (\ref{147}) as
\be S^{\,_A}_{_{\rm F}\,_B}=H^{_{AC}}F_{_{CB}}+
\Euro_{_{\rm F}}\gamma^{_A}_{_B}\, .\label{150}\fe
which is equivalent to the expression given, using the notation
(\ref{124}), as
\be S_{_{\rm F}}^{\,\mu\nu}={1\over 8\pi}\big(2 B^\mu B^\nu-B^2
\gamma^{\mu\nu}\big)\, .\label{151}\fe

In order to consider the effect on wave propagation, as discussed in
Section 8, we need the ensuing replacement of (\ref{115a}) for the tensor
defined in the relativistic case by (\ref{110a}), which works out as
{\be y^{\,\mu\nu}=\left(1+\frac{\Euro_0+P_0}{\rho\, c^2}\right)
\gamma^{\mu\nu}+\frac{1}{4\pi c^2\rho}(B^2\gamma^{\mu\nu}- B^\mu B^\nu)
\, .\label{151a}\fe}
We also need to evaluate the corresponding (unpolarised) magnetic elasticity
contribution. This will be obtainable on the basis of the ansatz
(\ref{101}), which provides the expression
$$ A_{_{\rm F}}^{\, _{ACBD}}= E_{_{\rm F}}^{\, _{ACBD}}+ \gamma^{_{AB}} 
S_{_{\rm F}}^{\,_{CD}}=H^{_{AB}}B^{_{CD}}+H^{_{DA}}B^{_{BC}}+\Euro_{_{\rm F}}
(\gamma^{_{AB}}\gamma^{_{CD}}+2\gamma^{_A[_D} \gamma^{_C]_B})$$
\be \hskip 3 cm +2 S_{_{\rm F}}^{\,_A[_C}\gamma^{_D]_B}+2
S_{_{\rm F}}^{\,_B[_C}\gamma^{_D]_A}- S_{_{\rm F}}^{\,_{AB}}
\gamma^{_{CD}}\, .\label{152}\fe
It immediately follows that the corresponding  magnetic contribution in the 
characteristic matrix $Q^{\mu\nu}$ given by (\ref{114}) will be expressible 
in terms of the relevant magnetic field tensor as 
\be Q_{_{\rm F}}^{\, \mu\nu}=A_{_{\rm F}}^{\, \mu\rho\nu\sigma}\nu_\rho
\nu_\sigma=-4\pi( H^{\mu\rho}H_\rho^{\ \, \nu} +
H^{\mu\rho}H^{\nu\sigma}\nu_\rho\nu_\sigma)\, .\label{153}\fe
By introducing a spacelike unit vector $\iota^\mu$ that is chosen in the
polarisation plane orthogonal to $\nu^\mu$ (the unit vector in the direction
of polarisation) in such a way that the magnetic induction vector will be 
expressible as
\be B^\mu={_1\over ^2}\epsilon^{\mu\nu\rho}F_{\nu\rho}=B_{_\Vert} \nu^\mu
+B_\perp \iota^\mu\, ,\label{154}\fe 
we can rewrite (\ref{153}) as
 \be 4\pi Q_{_{\rm F}}^{\, \mu\nu}=B_\perp^2\nu^\mu\nu^\nu-2B_{_\Vert} 
B_\perp \nu^{(\mu} \iota^{\nu)}+B_{_\Vert}^2(\gamma^{\mu\nu}
-\nu^\mu\nu^\nu)\, .\label{155}\fe

The complete characteristic matrix,
\be Q^{\mu\nu}=Q_0^{\,\mu\nu}+Q_{_{\rm F}}^{\,\mu\nu}\, \label{156}\fe
can easily be obtained in an explicit form if we suppose that the material 
contribution $Q_0^{\,\mu\nu}$ has the form (\ref{118}) that is relevant 
when the medium is in a simple isotropic state of the kind characterised 
by (\ref{116}). In this (unpolarised, materially isotropic) case, we 
obtain  an expression of the form
\be Q^{\mu\nu}=Q_{_\Vert}\nu^\mu\nu^\nu-2Q_\times
 \nu^{(\mu} \iota^{\nu)}+Q_\perp(\gamma^{\mu\nu}-\nu^\mu\nu^\nu)
\, .\label{157}\fe
with
\be Q_{_\Vert}=\beta_0 +{4\mu_0\over 3}+{B_\perp^2\over 4\pi}\, ,\hskip
1 cm Q_\times={B_{_\Vert}B_\perp\over 4\pi}\, ,\hskip 1 cm
Q_\perp=\mu_0+{B_{_\Vert}^2\over 4\pi}\, .\label{158}\fe

The characteristic eigenvector equation (\ref{112}) is easily soluble
in the non-relativistic case,  for which we simply have 
$y^{\mu\nu}=\gamma^{\mu\nu}$: it can be seen that there will always be a 
transverse Alfven type mode, with polarisation covector $\hat u_\mu$ 
orthogonal to both the propagation direction and the magnetic field 
direction, whose velocity $\upsilon_{_\top}$ will be given by
\be \upsilon_{_\top}^2={Q_\perp\over\rho}={\mu_0\over\rho}+{B_{_\Vert}^2
\over 4\pi\rho}\, ,\label{159}\fe
an expression that reduces to the well known formula
$\upsilon_{_\top}^2=\mu_0/\rho$ for propagation of transverse (``wiggle'' or
``shake'') modes in an isotropic solid when the magnetic field is absent.

In the case of propagation parallel to the magnetic field direction,
meaning $\nu^\mu\propto B^\mu$ so that $B_\perp=0$, there will be a 
second (orthogonally polarised) transverse mode, with the same propagation 
speed $\upsilon_{_\top}$, which in this case will be given by the expression 
$\upsilon_{_\top}^2=\mu_0/\rho+B^2/4\pi\rho$, which reduces to the well known 
Alfven formula $\upsilon_{_\top}^2=B^2/4\pi\rho$ in the magnetohydrodynamic 
(purely fluid) limit in which the rigdity coefficient $\mu$ vanishes. There 
will also be a purely longitudinal (sound type) mode with velocity 
$\upsilon_{_\Vert}$ that will be given, independently of the magnetic field 
strength, by the expression
\be \upsilon_{_\Vert}^2=\frac{\beta_0}{\rho}+\frac{4\mu_0}{ 3\rho}
\, ,\label{160}\fe
which reduces in the magnetohydrodynamic limit, $\mu\rightarrow 0$, just 
to Newton's formula $v_{_\Vert}^2={\rm d}P/{\rm d}\rho$ for the speed of 
ordinary sound.

More generally, when the propagation is not parallel to the magnetic
field, the other two modes will  be of mixed -- partially longitudinal, 
partially transverse -- type with polarisation in the plane generated by 
the propagation direction and the magnetic field direction, and with 
speeds $\upsilon_+$ and $\upsilon_-$ that will be obtainable as the roots 
of the eigenvalue equation
\be (Q_{_\Vert}-\rho \upsilon^2)(Q_\perp-\rho \upsilon^2)-Q_\times^2=0
\, ,\label{161}\fe
which gives
\be 2\rho\,\upsilon_\pm^2=Q_{_\Vert}+Q_\perp\pm\sqrt{(Q_{_\vert}-
Q_\perp)^2+4Q_\times^2}\, .\label{162}\fe
These solutions can be seen to be such that we shall have
$\upsilon_+\rightarrow\upsilon_{_\Vert}$ and $\upsilon_-
\rightarrow \upsilon_{_\top}$ in the limit of parallel propagation for 
which $B_\perp\rightarrow 0$ and $B_{_\Vert}\rightarrow B$.

In the relativistic case we shall still be able to use the same expressions 
(\ref{157}) and (\ref{158}) for $Q^{\mu\nu}$, but for the tensor 
$y^{\mu\nu}$ in the characteristic equation (\ref{112}) it will be
necessary to use the less simple formula (\ref{151a}), which will be
expressible, in a form analogous to (\ref{157}), as
{\be \rho\, c^2 y^{\mu\nu}=
Y_{_\Vert}\nu^\mu\nu^\nu-2Q_\times
 \nu^{(\mu} \iota^{\nu)}+Y_\perp\iota^\mu\iota^\nu+ Y
(\gamma^{\mu\nu}-\nu^\mu\nu^\nu-\iota^\mu\iota^\nu)
\, .\label{157a}\fe}
with
{\be Y_{_\Vert}=c^2\tilde\rho_{\,0}+\frac{B_\perp^2}{4\pi}\, ,\hskip 1cm
Y{_\perp}=c^2\tilde\rho_{\,0}+\frac{B_{_\Vert}^2}{4\pi}\, ,\hskip 1cm
Y=c^2\widetilde\rho_{\,0}+\frac{B^2}{4\pi}\, ,\label{158a}\fe}
where
{\be c^2\widetilde\rho_{\,0}=c^2\rho+\Euro_0+P_0\, .\label{163}\fe}

As before, there will always be a transverse Alfven type mode, with 
polarisation covector $\hat u_\mu$ orthogonal to both the propagation 
direction and the the magnetic field direction, with velocity 
$\upsilon_{_\top}$ that will be given generically by
{\be \frac{\upsilon_{_\top}^2} {c^2}= \frac{Q_\perp}{Y}=\frac{4\pi\mu_0
+B_\Vert^2}{4\pi c^2\widetilde\rho_{\,0}+B^2}\, .\label{159a}\fe}
In the case of propagation in the direction of the magnetic field,
i.e. when $B_\perp=0$ there will again be a second (orthogonally
polarised) transverse mode, with the same propagation 
speed $\upsilon_{_\top}$, as well as a purely longitudinal (sound type)
 mode with velocity $\upsilon_{_\Vert}$ that will be given in the
generic case by the same formula as has long been well known~\cite{C73} 
for  the unmagnetised case, namely
{\be \upsilon_{_\Vert}^2=\frac
{\beta}{\widetilde\rho_{\,0}}+\frac{4\mu_0}{ 3\widetilde\rho_{\,0}}
\, .\label{160a}\fe}

For the generic case, in which  the propagation is not parallel to the 
magnetic field, the  relativistic generalisation of the equation
(\ref{162}) for the speeds  $\upsilon_+$ and $\upsilon_-$ of
the mixed modes -- with polarisation in the plane generated by 
the propagation direction and the magnetic field direction --
will be expressible in terms of the dimensionless ratios
{\be {\cal Q}_{_\Vert}=\frac{Q_{_\Vert}}{Y_{_\Vert}}\, ,\hskip 1 cm
 {\cal Q}_{\perp}=\frac{Q_{\perp}}{Y_{\perp}}\, ,\hskip 1 cm
{\cal Q}_{\times}=\frac{ Q_\times}{\sqrt{Y_{_\Vert} Y_\perp}}
\, ,\label{164}\fe}
by 
{\be \frac{\upsilon_\pm^2} {c^2}=\frac{  {\cal Q}_{_\Vert}+ 
{\cal Q}_{\perp}-2{\cal Q}_{\times}^{\,2} \pm\sqrt{({\cal Q}_{_\Vert}
-{\cal Q}_\perp)^2+4{\cal Q}_\times^{\,2}(1-{\cal Q}_{_\Vert})
(1-{\cal Q}_\perp)}}{2(1-{\cal Q}_{\times}^{\,2})}\, ,\label{162a}\fe}

\bigskip
{\bf Acknowledgements.} The authors wish to thank S. Bonazzola for
stimulating conversations. Nicolas Chamel acknowledges support
from the Lavoisier program of the French Ministry of Foreign Affairs.

\vfill\eject


\medskip
\begin{thebibliography}{99}

\bibitem{CK92} B. Carter \& I.M. Khalatnikov,
``Momentum, Vorticity, and Helicity in Covariant Superfluid Dynamics",
{\it Ann. Phys.} {\bf 219} (1992) 243-265.

\bibitem{LSC97}  D. Langlois, D. Sedrakian, B.Carter,
``Differential rotation of relativistic superfluid in neutron stars'',
{\it Mon. Not. R. Astr. Soc.}, {\bf 297} (1998)  1198-1201.
[astro-ph/9711042]

\bibitem{Kunzle86} H.P. Kunzle,
``Lagrangian formalism for adiabatic fluids on five-dimensional
space-time'',
{\it Canad. J. Phys.} {\bf 64} (1986) 185-189

\bibitem{DuvalGibbonsHorvathy91} C. Duval, G. Gibbons, P. Horvathy,
``Celestial mechanics, conformal structures, and gravitational waves'',
{\it Phys. Rev.} {\bf D43} (1991) 3907-3902.

\bibitem{CK94} B. Carter \& I.M. Khalatnikov,
``Canonically covariant formulation of Landau's Newtonian superfluid
dynamic'',
{\it Rev. Math. Phys.} {\bf 6} (1994)  277-304.

\bibitem{Cartan23} E. Cartan,
``Sur les variétés à connexion affine et la théorie de la relativité généralisée'',
{\it Ann. Sci. Ecole Norm. Sup.} {\bf 40} (1923) 325-412 ;
{\it Ann. Sci. Ecole Norm. Sup.} {\bf 41} (1924) 1-25 ;
{\it Ann. Sci. Ecole Norm. Sup.} {\bf 42} (1925) 17-88.

\bibitem{CC02} B. Carter, N. Chamel
``Covariant analysis of Newtonian multi-fluid models for neutron stars:
I Milne - Cartan structure and variational formulation''
{\it Int. J. Mod. Phys.} {\bf D13} (2004) 291-325.
[astro-ph/0305186]

\bibitem{CC03} B. Carter, N. Chamel
``Covariant analysis of Newtonian multi-fluid models for neutron stars:
II Stress - energy tensors and virial theorems'', to appear in
{\it Int. J. Mod. Phys.} (2005).
[astro-ph/0312414]

\bibitem{CC04} B. Carter, N. Chamel
``Covariant analysis of Newtonian multi-fluid models for neutron stars:
III Transvective, viscous, and superfluid drag dissipation'', to appear in
{\it Int. J. Mod. Phys.} {\bf D} (2005).
[astro-ph/0410660]

\bibitem{BCCM05} R.A. Battye, B. Carter, E. Chachoua, A. Moss,
``Rigidity and stability of cold dark solid universe model'',
 to appear in {\it Phys. Rev.} {\bf D} (2005).
[hep-th/0501244]

 \bibitem{Palmer2005} D.M.Palmer \textit{et al.},
 `` A Giant $\gamma$-ray flare from the magnetar SGR 1806-20'',
 {\it Nature } {\bf 434} (28 Apr 2005)  1107-1109.
[astro-ph/0503030]


\bibitem{Eichler02} D. Eichler, 
``Waiting for the big one: a new class of soft gamma ray repeater outbursts'',
{\it Mon. Not. R. Astr. Soc.}, {\bf 335} (2002) 883-886.
[astro-ph/0204512]

 \bibitem{Hurley05} K.Hurley \textit{et al.},
 `` An exceptionally bright flare from SGR 1806-20 and the origins of 
short-duration $\gamma$-ray bursts'',
 {\it Nature } {\bf 434} (28 Apr 2005)  1098-1103.
[astro-ph/0502329]

\bibitem{Trautman65} A. Trautman,
``Invariance properties and conservation laws''
in {\it Brandeis Lectures on General Relativity}
ed. A. Trautman, F.A.E. Pirani, H. Bondi
(Prentice Hall, New Jersey, 1965) 158 - 200.

\bibitem{C73} B. Carter,
``Elastic Perturbation Theory in General Relativity and a Variational
Principle for a Rotating Solid Star",
{\it Commun. Math. Phys.}  {\bf 30} (1973) 261-286.

\bibitem{C89} B. Carter,
``Covariant Theory of Conductivity in Ideal Fluid or Solid Media",
B. Carter, in {\it Relativistic Fluid Dynamics (C.I.M.E., Noto, May 1987)}
ed.  A.M. Anile, \& Y. Choquet-Bruhat, Lecture Notes in Mathematics {\bf 1385}
(Springer - Verlag, Heidelberg, 1989) 1-64.

\bibitem{Schutz} B. Schutz,
``Perfect Fluids in General Relativity: Velocity Potentials and a
Variational Principle'',
{\it Phys. Rev.} {\bf D2} (1970) 2762-2773.

\bibitem{LL} L.D. Landau, E.M. Lifshitz,
{\it Theory of Elasticity} (Pergamon, London, 1959).

\bibitem{Souriau65} J.M. Souriau,
{\it G\'eometrie et Relativit\'e} (Herman, Paris, 1965)

\bibitem{DeWitt62} B. DeWitt,
``The quantisation of geometry''
in {\it Gravitation:an introduction to current research},
ed L. Witten (Wiley, New York, 1962) 266-381.

\bibitem{CQ72} B. Carter, H. Quintana,
``Foundations of General Relativistic High Pressure Elasticity Theory",
{\it Proc. Roy. Soc. Lond.} {\bf A331} (1972) 57-83.

\bibitem{C83} B. Carter,
``Interaction of Gravitational Waves with an Elastic Solid Medium",
in {\it Gravitational Radiation} (Proc. 1982 Les Houches Summer School),
ed N. Deruelle \& T. Piran, (North Holland, Amsterdam, 1983) 455-464.
[gr-qc/0102113]


\bibitem{C94} B. Carter,
``Kalb-Ramond Vorticity Variational Formulation of Relativistic
Perfectly Conducting Fluid Theory",
{\it Int. J. Mod. Phys.} {\bf D3} (1974) 15-21.

 \bibitem{C99} B. Carter,
 ``Vortex dynamics in superfluids'',  in
{\it Topological defects and non-equilibrium dynamics of symmetry breaking
 phase transitions} (NATO ASI {\bf C549}, Les Houches, February 1999)
ed Yu.M. Bunkov, H. Godfrin,  (Kluwer, Dordrecht, 2000) 267 - 301.
[gr-qc/9907039]

\bibitem{C72} B. Carter,
``Speed of sound in a high pressure general relativistic solid''
{\it Phys. Rev.} {\bf D7} (1973) 1590-1593.

\bibitem{CG88} B. Carter, B. Gaffet,
``Standard covariant formulation for perfect fluid dynamics'',
{\it J. Fluid. Mech.} {\bf 186} (1988) 1-24.

\bibitem{BekOron00} J.D. Bekenstein, A. Oron
``Conservation of circulation in magnetohydrodynamics''
{\it Phys. Rev.}  {\bf E62} (2000) 5594-5603.
[astro-ph/0002045]

\bibitem{BekOron78} J. D. Bekenstein, E. Oron
``New conservation laws in general-relativistic magnetohydrodynamics''
{\it Phys. Rev.} {\bf D18} (1978) 1809-1819.

\bibitem{Woltjer58} L. Woltjer,
``A theorem on force-free magnetic fields'',
{\it Proc. Nat. Acad. Sci. USA} {\bf 44} (1958) 489-491.


 
\end{thebibliography}
\end{document}